\documentclass[twocolumn,showpacs]{revtex4-1}

\usepackage{color}
\usepackage{graphicx}

\begin{document}  

\title{Orbital mixing in few-layer graphene and non-Abelian Berry phase
}
\author{K. Shizuya}
\affiliation{Yukawa Institute for Theoretical Physics\\
Kyoto University,~Kyoto 606-8502,~Japan }

\begin{abstract}

In a magnetic field few-layer graphene supports, at the lowest Landau level, 
a multiplet of zero-mode levels nearly degenerate in orbitals as well as in spins and valleys.
Those pseudo-zero-mode (PZM) levels are generally sensitive to interactions and  external perturbations,
and have a crossing among themselves or with other higher Landau levels when an external field 
is swept over a certain range.
A close study is made of how such PZM levels evolve when they are gradually brought 
from empty to filled levels under many-body interactions. 
It is pointed out that the level spectra generally avoid a crossing via orbital level mixing 
and that orbital mixing is governed by a non-Abelian Berry phase 
that derives from an approximate degeneracy and interactions.
A look is also taken into evolution/crossing of many-body ground states 
with increasing external bias in bilayer graphene. 

\end{abstract} 


\maketitle

\section{Introduction}

Graphene hosts massless Dirac electrons as charge carriers that display fascinating electronic properties.
Recently considerable attention centers on graphene bilayers and few-layers~\cite{NMMKF,OBSHR,MF,GCP,KA}, 
where the added layer degrees of freedom open a new realm of physics and applications 
with, e.g., a tunable band gap~\cite{MF,OBSHR}  in bilayer graphene.

In a magnetic field few-layer graphene supports, at the lowest Landau level (LLL), 
a multiplet of  zero-energy levels degenerate in Landau orbitals 
as well as in spins and valleys.
Bilayer graphene supports an octet~\cite{MF} $(2_{\rm spin}\times 2_{\rm valley}\times 2)$ 
of such levels with a two-fold degeneracy in Landau orbitals $n = \{0,1\}$. 
Trilayers acquire a three-fold degeneracy in orbitals.
This orbital degeneracy has a topological origin in the index of (the leading part of) the one-body Dirac Hamiltonian. 
In the presence of spin and band anisotropies and many-body interactions, 
these zero-mode levels evolve into pseudo-zero-mode (PZM) levels, 
or into a variety of broken-symmetry quantum Hall states, 
as discussed theoretically~\cite{BCNM,KSpzm,BCLM,CLBM,CLPBM,NL,GGJ,KS_Ls,Khari,LC,KJ,KScrBG} 
and explored experimentally~\cite{FMY,ZCZJ,WAFM,MFW,KLT,VJB,MDY,LFX,HLZW}. 
It was noted, in particular, that the orbital degeneracy is also lifted by 
Coulomb interactions alone~\cite{KS_Ls}, 
with the zero-energy modes  orbitally Lamb shifted 
due to quantum fluctuations of the filled valence band. 
The orbital degeneracy and its lifting by the orbital Lamb shift 
are new features specific to the LLL in few-layer graphene.

Those PZM levels are generally sensitive to external perturbations
and have a chance of crossing among themselves or with other higher Landau levels 
when an external field is swept over a certain range.
Many-body interactions significantly affect such level-crossing phenomena.
To see this, let us suppose that 
two empty levels, that differ only in orbitals $n$ and $m$ ($n>m\ge0)$,  have a crossing 
when an external field $u$ is varied across a critical value $u^{\rm cr}$, 
as depicted in Fig.~1(a), and ask what happens when one fills them with electrons. 
If there is no electron-electron interaction, the level spectra remain unchanged
and the filled levels continue to have a crossing at $u^{\rm cr}$.

 \begin{figure}[tbp]
\begin{center}
\includegraphics[scale=0.7]{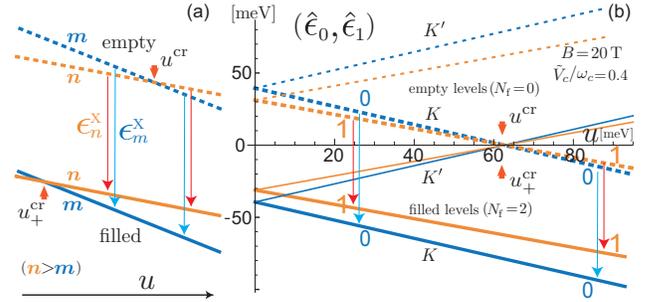}
\end{center}
\vskip-.7cm
\caption{Orbital level crossing. 
(a) Two empty levels (dotted lines) of different orbitals $m$ and $n$ undergo a crossing 
with increasing  external bias $u$. 
When they are filled up (solid lines), 
the crossing point gets shifted ($u^{\rm cr} \rightarrow u^{\rm cr}_{+})$
due to orbit-dependent Coulombic exchange energies, causing
a level inversion for $u \in (u^{\rm cr}_{+}, u^{\rm cr})$.
(b)~Renormalized PZM level spectra $(\hat{\epsilon}_{0}, \hat{\epsilon}_{1})$ in bilayer graphene,
plotted as a function of  external bias $u$. 
A level inversion takes place for $0\le u< u^{\rm cr}$ in valley $K$ and for $-u^{\rm cr} < u\le  0$ in valley $K'$. }
\end{figure}

In the presence of interaction, the level spectra are lowered by the amount of exchange energies, 
and the filled levels lose a crossing at $u \sim u^{\rm cr}$, or the crossing point gets shifted, 
$u^{\rm cr} \rightarrow u^{\rm cr}_{+}$, as illustrated in the figure. 
(Note that the exchange energy $\epsilon^{\rm x}_{n} < 0$ generally decreases 
with increasing orbital index $n$;
$|\epsilon^{\rm x}_{n}| < |\epsilon^{\rm x}_{m}|$ for $n>m \ge 0$.)
The two levels, as they are gradually filled, 
thus appear to cross for $u \in  (u^{\rm cr}_{+},  u^{\rm cr})$
while no crossing is expected for $u > u^{\rm cr}$.

Experimentally a similar many-body phenomenon of spin exchange energy origin has been known~\cite{ZFJ}:  
In a GaAs/AlGaAs quantum well with doubly occupied subbands, crossings of  two Landau levels of different subbands lead to ring-like structures in the phase diagram that suggest transitions due to spin exchange energy.

The purpose of the present paper is to examine how those nearly degenerate PZM levels 
behave when they are gradually brought from empty to filled levels under many-body interactions. 
It is pointed out that the level spectra generally avoid a crossing via orbital level mixing and 
that orbital mixing is governed by a non-Abelian Berry phase~\cite{BerryPh,WZ} 
that derives from an approximate degeneracy and interactions.
This non-Abelian phase clarifies the algebraic features underlying 
the phenomena of Landau-level crossing and mixing.  
We also examine, as a typical case of crossing of many-body states, 
how the neutral ($\nu=0$) ground state in bilayer graphene evolves 
with increasing interlayer bias $u$.

In Sec.~II we briefly review some basic features of the PZM levels in bilayer graphene.
In Secs.~III and IV,  we examine the level-mixing phenomenon and its algebraic character 
in the light of a non-Abelian Berry phase.  
In Sec.~V we look into trilayer graphene and show that the Berry phase encodes and 
distinguishes possible patterns of orbital mixing in different types of trilayers. 
In Sec.~VI we examine evolution of the $\nu=0$ ground state in bilayer graphene with bias $u$. 
Section VII is devoted to a summary and discussion.

\section{The lowest Landau level in bilayer graphene}

In a magnetic field $B_{z}= B$  one-body states  $|n,y_{0};a,\alpha\rangle$ 
of a Dirac electron in graphene are labelled by integers $n=0,\pm1, \pm 2, \cdots$ 
and momentum $p_{x}$ (or the center-coordinate $y_{0} \equiv \ell^2 p_{x}$ 
with the magnetic length $\ell \equiv1/\sqrt{eB}$), as well as valleys $a \in (K,K')$
and spins $\alpha\in (\uparrow, \downarrow)$.
The associated one-body Hamiltonian is generally written as 
\begin{equation}
H^{\rm 1b} = \int\! dy_{0}\, \sum_{n,a,\alpha} 
{\psi^{n,a }_{\alpha}}^{\dag}(y_{0})\,  \epsilon_{n}^{a;\alpha} \psi^{n,a}_{\alpha}(y_{0}),
\label{H_onebody}
\end{equation}
where $\psi^{n,a}_{\alpha}(y_{0})$ denotes the electron field 
with the spectrum $\epsilon_{n}^{a;\alpha}$;
$\psi^{n,a}_{\alpha}(y_{0}) \equiv \int d^{2}{\bf x}\, \langle n,y_{0};a, \alpha|{\bf x}\rangle \Psi_{\bf x}$ 
in terms of the field $\Psi_{\bf x}$ in  the coordinate space. 
The charge density 
$\rho_{-{\bf p}} =\int d^{2}{\bf x}\,  e^{i {\bf p\cdot x}}\, \Psi_{\bf x}^{\dag}\Psi_{\bf x}$
is thereby written as
\begin{eqnarray}
\rho_{-{\bf p}} &=& \gamma_{\bf p}\sum_{m, n =-\infty}^{\infty}
\sum_{a,\alpha} g^{m n;a}_{\bf p}\, 
R^{m n;aa}_{\alpha\alpha;{\bf -p}}, 
\nonumber\\
R^{m n;ab}_{\alpha\beta;{\bf -p}}&\equiv& \int dy_{0}\,
{\psi^{m,a}_{\alpha}}^{\dag}(y_{0})\, e^{i{\bf p\cdot r}}\,
\psi^{n,b}_{\beta} (y_{0}),
\label{chargeoperator}
\end{eqnarray}
with  $\gamma_{\bf p} \equiv  e^{- \ell^{2} {\bf p}^{2}/4}$;
${\bf r} = (i\ell^{2}\partial/\partial y_{0}, y_{0})$
stands for the center coordinate with uncertainty 
$[r_{x}, r_{y}] =i\ell^{2}$.
The projected charges $R^{mn;ab}_{\alpha\beta;{\bf p}}$ obey the $W_{\infty}$ algebra~\cite{GMP}.
The coefficient functions $g^{mn;a}_{\rm p}$ have the structure 
\begin{equation}
g^{nm;a}_{\bf p} \propto p^{|m| - |n|} \times ({\rm polynomials\ of}\ \ell^2 {\bf p}^2)\ 
{\rm for}\ |m|\ge |n|, 
\end{equation}
and $g^{mn;a}_{\bf p}= (g^{nm;a}_{\bf -p})^{\dag}$, with $p\equiv p_{x} + ip_{y}$.

For bilayer graphene the one-body spectra $\{\epsilon_{n}^{a;\alpha}\}$ 
take an electron-hole ($e$-$h$) symmetric pattern
when only the leading intralayer and interlayer couplings 
$\gamma_{0} \equiv \gamma_{AB} \sim 3\,$eV (related to the Fermi velocity 
$v \sim 10^{6}$m/s in monolayer graphene) and 
$\gamma_{1} \equiv \gamma_{A'B} \sim 0.4\,$eV are kept. 
For simplicity, nonleading couplings $(\Delta, \gamma_{4}, \cdots)$ 
that lead to  weak $e$-$h$ breaking~\cite{ZLBF,LHJ_asym} 
are suppressed in what follows;  spin splitting is to be restored in Sec.~VI.

The Coulomb interaction is written as 
$V^{C} ={1\over{2}} \sum_{\bf p} v_{\bf p}\,  {:\! \rho_{\bf -p}\, \rho_{\bf p}\!:}$,
with potential 
$v_{\bf p}= 2\pi \alpha_{e}/(\epsilon_{\rm b} |{\bf p}|)$,
$\alpha_{e} \equiv e^{2}/(4 \pi \epsilon_{0})$ and 
the substrate dielectric constant $\epsilon_{\rm b}$; $\sum_{\bf p} \equiv \int d^{2}{\bf p}/(2\pi)^{2}$
and $:\ :$ denotes normal ordering.
For simplicity, we ignore a tiny interlayer separation $d \rightarrow 0$.   
It is advantageous to cast $V^{C}$ 
in the form of manifest exchange interaction,
\begin{equation}
V^{C} = -  {1\over{2}} \sum_{\bf k} \tilde{v}_{\bf k}^{jk;mn;ab} 
:{R}^{mk;ba}_{\beta\alpha; {\bf -k}} {R}^{jn;ab}_{\alpha\beta; {\bf k}}: ,
\label{VCdual}
\end{equation}
with the dual potential
\begin{equation}
\tilde{v}^{jk;mn;ab}_{\bf k}=  {1\over{\bar{\rho}}}\sum_{\bf p} v_{\bf p}\gamma_{\bf p}^{2}\,
g^{jk;a}_{\bf p}\, g^{mn;b}_{\bf -p} e^{i \ell^{2}  {\bf p \times k}},
\end{equation}
where $\bar{\rho} = 1/(2\pi \ell^2)$ and ${\bf p \times k} = p_{x}k_{y}- p_{y}k_{x}$.
(For clarity, summation $\sum$ over repeated labels will be suppressed from now on.)
This direct-exchange duality of the Coulomb interaction is made manifest 
on the operator level~\cite{ks_duality}
for planar electrons in a magnetic field, 
where interaction becomes short-ranged with a cutoff $\sim \ell$.

In bilayer graphene an octet of PZM levels, 
nearly degenerate in spins, valleys and orbitals $n=\{ 0,1\}$,
forms the LLL isolated from other Landau levels.
A key feature is a band gap which is 
tunable~\cite{MF,OBSHR} by an applied interlayer bias $u$.
Actually, bias $u$ splits valleys $(K, K')$ and 
the bare spectra have the following valley structure,
\begin{equation}
\epsilon_{-n}^{K} = -\epsilon_{n}^{K}|_{-u}, \ \ \epsilon_{n}^{K'} = \epsilon_{n}^{K}|_{-u},
\label{bias_valley}
\end{equation}
where ${\cal O}|_{-u}$ signifies setting $u\rightarrow -u$ in ${\cal O}$.
Here each $n=\pm 2, \pm 3,\dots$  
refers to a pair of electron and hole levels. 
In contrast, the PZM levels $n=\{0,1\}$ stand alone (per spin and valley) and
are $e$-$h$ self-conjugate, with $\pm n \rightarrow n$ in Eq.~(\ref{bias_valley}).
Their spectra read~\cite{KScrBG} 
\begin{eqnarray}
\epsilon_{0}^{K} &=& - u/2,
 \ \  \epsilon_{1}^{K} = - z_{1}\, u/2,
\\
z_{1} &=& 1 - 2/(g^2 +1) + O(u^2/g^{6}\omega_{c}^{2}) <1,
\end{eqnarray}
where $g \equiv \gamma_{1}/\omega_{c}$ and 
\begin{equation}
\omega_{c} \equiv \sqrt{2}\, v/\ell \approx 36.3 \times v[10^{6}{\rm m/s}] \sqrt{B[T]}\, {\rm meV}
\end{equation}
is the  characteristic cyclotron energy of graphene.
Note that $\epsilon_{1}$ has a slightly smaller gradient ($z_{1}<1$) in bias $u$ than $\epsilon_{0}$.
As for band parameters~\cite{JM}, we adopt 
$v=0.845 \times 10^{6}$m/s and $\gamma_{1} = 361$\,meV so that 
$\omega_{c} \approx 137$ meV, $g \approx 2.63$ and 
$z_{1} \approx 0.75$ at $B=20$\,T.

Interlayer bias $u$ shifts the PZM levels $n=\{0,1\}$ oppositely $(\propto \mp u/2)$ in the two valleys.  
We take, without loss of generality, $u \ge 0$ for valley $K$; $u < 0$ then refers to $K'$.
The valley gap $\sim u$ increases with bias $u$
while  $(\epsilon_{0},\epsilon_{1})$ remain nearly degenerate in each valley.

For the PZM levels, 
form factors $g^{m,n;a}_{\bf p}$ take particularly simple form   
\begin{eqnarray}
g^{00}_{\bf p} &=& 1, \ \
g^{11}_{\bf p} =1-c_{1}^{2}\, \textstyle{1\over{2}}\ell^{2}{\bf p}^{2}, 
\nonumber\\
g^{01}_{\bf p} &=& c_{1} \ell\, p/\sqrt{2},\ \ 
g^{10}_{\bf p} = -c_{1} \ell\, p^{\dag}/\sqrt{2},
\label{gmnPZM}
\end{eqnarray}
with $c_{1}|^{K'} = c_{1}|^{K}_{- u}$; in $e$-$h$ symmetric setting, 
\begin{equation}
c_{1} \approx   1/\sqrt{1 + (1/g^2)} \stackrel{B=20{\rm T}}{\sim} 0.93
\end{equation}
scarcely depends on bias $u$ and valleys.

Electrons in each Landau level are subject to Coulombic quantum fluctuations of the filled valence band.
The exchange interaction gives rise to $O(V^{C})$ self-energy corrections 
to level spectra $\epsilon_{n}^{a; \alpha}$ of the form
\begin{equation}
\delta \epsilon_{n}^{a; \alpha}
= -\sum_{\bf p}v_{\bf p}\gamma_{\bf p}^{2}\,
\sum_{m} \nu_{m}^{a; \alpha}\, |g^{n m;a}_{\bf p}|^2, 
\label{self_En}
\end{equation}
where $0 \le \nu_{m}^{a; \alpha} \le 1$ stands for the filling fraction 
of the $(m, a, \alpha)$ level. 
The direct interaction leads to corrections $\propto v_{\bf p \rightarrow 0}$, which, as usual, 
are removed when the neutralizing background is assumed. 
The exchange energy acts separately for each (valley, spin) channel. 
We thus suppress those labels below and mainly refer to valley $K$. 
We use $0\le N_{\rm f} \le 2$ to specify the filling fraction of the PZM sector $n= \{0,1\}$ (per spin and valley), 
with $N_{\rm f}=0$ for the empty sector and $N_{\rm f}=2$ for the filled one.

Let us now consider an empty PZM sector with levels below it all filled, 
i.e., $\nu_{n}^{a;\alpha} = 1$ for $n \le -2$.
Infinitely many filled levels in the valence band make self-energies $\delta \epsilon_{n}^{a; \alpha}$
ultraviolet divergent, and one has to go through renormalization of band parameters $v$ and $\gamma_{1}$.
See Ref.~\cite{KScrBG} for details of the renormalization procedure.
The empty PZM levels in valley $K$, e.g., acquire the following {\em renormalized} spectra
\begin{eqnarray}
\hat{\epsilon}_{0}|^{N_{\rm f}=0} 
&=& - {\textstyle{1\over{2}}}u+ \Omega_{0} 
+ {\textstyle{1\over{2}}} ( 1+  {\textstyle{1\over{2}}} c_{1}^2 )\, \tilde{V}_{c},
\nonumber\\
\hat{\epsilon}_{1}|^{N_{\rm f}=0}
&=& - {\textstyle{1\over{2}}} z_{1} u +\Omega_{1}
+ {\textstyle{1\over{2}}} ( 1+ {\textstyle{1\over{2}}} c_{1}^2 
-  {\textstyle{1\over{4}}} C)  \tilde{V}_{c}, \ \
\label{Spec_pzm_empty}
\end{eqnarray}
with $C= (4 - 3  c_{1}^{2})\, c_{1}^{2} \stackrel{B=20{\rm T}}{\sim} 1.20$ and
\begin{equation}
\tilde{V}_{c} \equiv \sum_{\bf p}v_{\bf p}\gamma_{\bf p}^{2}
= {\alpha_{e}\over{\epsilon_{b}\, \ell}}  \sqrt{{\pi\over{2}}}
 \approx {70.3 \over{\epsilon_{b}}}\, 
\sqrt{B[{\rm T}]}\, {\rm meV};
\end{equation}
$\hat{\epsilon}_{n}|^{K'} = \hat{\epsilon}_{n}|^{K}_{- u}$. 
Here $\{\Omega_{n}\}$, with the property $\Omega_{-n}=- \Omega_{n}|_{-u}$, 
stand for corrections coming from filled levels deep in the valence band. 
For the PZM levels they are practically linear in $u$, 
with $(\Omega_{0}, \Omega_{1})  \stackrel{B=20{\rm T}}{\approx} 
(-0.646, -0.612) (\tilde{V}_{c}/\omega_{c})\, u/2$ in the present $e$-$h$ symmetric setting; 
for other levels  $\{\Omega_{n}\}$ are appreciable in magnitude even for $u\rightarrow 0$.

Note that interaction lifts the orbital degeneracy at zero bias $u=0$, 
with $\hat{\epsilon}_{1}$ getting lower than  $\hat{\epsilon}_{0}$ by
\begin{equation}
(\hat{\epsilon}_{0} - \hat{\epsilon}_{1})|_{u=0} 
= {\textstyle{1\over{8}} }\, C\,  \tilde{V}_{c}  \equiv \epsilon_{\rm Ls},
\end{equation}
i.e., the PZM levels get orbitally Lamb-shifted~\cite{KS_Ls}. 
Numerically,
$(\hat{\epsilon}_{0}, \hat{\epsilon}_{1}, \epsilon_{\rm Ls} )|_{u=0} 
\stackrel{B=20{\rm T}}{\approx} (0.72, 0.57, 0.15) \tilde{V}_{c}$.
Let us write, for $u\not=0$, the full $(0,1)$ shift as
\begin{equation}
\delta\epsilon =\hat{\epsilon}_{0}-\hat{\epsilon}_{1} 
\equiv (1 -\xi)\, \epsilon_{\rm Ls}, 
\end{equation}
with 
$\xi\equiv \{(1- z_{1})u/2+ \Omega_{1} -\Omega_{0}\}/\epsilon_{\rm Ls}
\approx u/(g^2 +1) \epsilon_{\rm Ls}$.
The spectra $(\hat{\epsilon}_{0},\hat{\epsilon}_{1})$ have a crossing 
at $\xi = u/u^{\rm cr} =1$ or across the critical bias 
\begin{equation}
u^{\rm cr} \approx  (g^2 +1)\, \epsilon_{\rm Ls} 
\stackrel{B= {\rm 20T}}{\sim} 1.2\,  \tilde{V}_{c}.
\end{equation}
Numerically, $(\epsilon_{\rm Ls}, u^{\rm cr})  \stackrel{B=20{\rm T}}{\approx}  (8.3, 62)\,$meV 
for the choice  $\tilde{V}_{c}/\omega_{c} =0.4$ or $\epsilon_{b} \approx 5.7$.

When the PZM levels are filled with electrons, the spectra get lower 
by the amount of exchange energy acting within the sector [see Eq.~(\ref{self_En})], 
\begin{equation}
(\hat{\epsilon}_{0},  \hat{\epsilon}_{1})|^{N_{\rm f}=2}
=  (\hat{\epsilon}_{0} - G^{00} -G^{01}, \hat{\epsilon}_{1} - G^{10} -G^{11}), 
\label{Spec_filled}
\end{equation}
where
\begin{equation}
G^{mn} = \sum_{\bf p} v_{\bf p} \gamma_{\bf p}^2\, |g^{mn}_{\bf p}|^2, \ 
G^{mm;nn} = \sum_{\bf p} v_{\bf p} \gamma_{\bf p}^2\, 
g^{mm}_{\bf p} g^{nn}_{\bf -p};
\label{Gmn}
\end{equation}
$G^{mn} =G^{nm}$.  (Here $G^{mm;nn}$ are defined for later use.)
Substituting the explicit values,
\begin{equation}
\{G^{00},G^{01},G^{11},G^{00;11}\}
= \textstyle  \{1,{1\over{2}} c_{1}^2,
 1- {1\over{4}} C, 1- {1\over{2}} c_{1}^2\} \tilde{V}_{c},
 \label{Gpzm}
\end{equation}
yields the spectra of the filled levels,
\begin{eqnarray}
\hat{\epsilon}_{0}|^{N_{\rm f}=2} \!
&=&\textstyle - {1\over{2}}u+ \Omega_{0} 
- {1\over{2}} ( 1+ {1\over{2}} c_{1}^2 )\, \tilde{V}_{c},
\nonumber\\
\hat{\epsilon}_{1}|^{N_{\rm f}=2}\!
&=&\textstyle - {1\over{2}}z_{1} u +\Omega_{1}
-{1\over{2}} ( 1+ {{1\over{2}}} c_{1}^2 - {1\over{4}} C)  \tilde{V}_{c}.\ \  
\label{Spec_pzm_full}
\end{eqnarray}
Here the  orbital Lamb shift is enhanced and reversed in sign, 
$(\hat{\epsilon}_{0}- \hat{\epsilon}_{1})|^{N_{\rm f}=2} =-(1 +\xi)\, \epsilon_{\rm Ls} <0$ 
for $0<\xi < 1$.
Actually, the filled and empty spectra are related  via $e$-$h$ conjugation [in Eq.~(\ref{bias_valley})],
\begin{equation}
(\hat{\epsilon}_{0}, \hat{\epsilon}_{1})|^{N_{\rm f}=2} 
= (-\hat{\epsilon}_{0}, -\hat{\epsilon}_{1})|^{N_{\rm f}=0}_{-u}.
\end{equation}
The {\it renormalized} PZM sector, when either {\em empty} or {\em filled}, 
becomes a unique eigenstate to $O(V^{C})$ of the total Hamiltonian $H^{\rm 1b}+ V^{C}$.

Figure~1(b) depicts a typical pattern of PZM spectra 
$(\hat{\epsilon}_{0}, \hat{\epsilon}_{1})|^{K+ K'}$  
(with $\hat{\epsilon}_{n}^{K'} =  \hat{\epsilon}_{n}^{K}|_{-u}$) per spin,
with a crossing  at $u=u^{\rm cr}$ for empty levels in valley $K$ and 
at $u=-u^{\rm cr}$ for filled ones in valley $K'$. 
The orbital Lamb shift, upon level filling, induces 
a level inversion  $(\hat{\epsilon}_{0} > \hat{\epsilon}_{1})|^{\rm empty}
 \rightarrow  (\hat{\epsilon}_{1} > \hat{\epsilon}_{0})|^{\rm filled}$ for
 bias $u \in (-u^{\rm cr}, u^{\rm cr})$.
[Actually,  when $e$-$h$ breaking due to $(\Delta, \gamma_{4}, \dots)$ is taken into account, 
$u^{\rm cr}$ becomes smaller 
 ($\sim$ 30\,meV at $B=20\, $T)  in valley $K$
and far larger ($\sim -100$ meV) in $K'$~\cite{KScrBG}.
Accordingly, we focus, in what follows, on quantum phenomena 
related to a crossing of empty levels in valley $K$.]
Such a level inversion signals a level crossing or instability with filling,
which actually is avoided via mixing of $n=\{0,1\}$ levels,
as noted earlier~\cite{KS_Ls}.

\section{Level mixing} 

In this section we refine an earlier analysis of orbital level mixing 
from a new angle.
We first note that the empty PZM sector $n=\{0,1\}$ (per valley and spin) to $O(V^{C})$ 
is described by the one-body Hamiltonian $H^{\rm 1b}$ of Eq.~(\ref{H_onebody}) 
with $(\epsilon_{0}, \epsilon_{1})$ replaced by the renormalized $(N_{\rm f}=0)$ spectra 
$(\hat{\epsilon}_{0}, \hat{\epsilon}_{1})$ in Eq.~(\ref{Spec_pzm_empty});
we denote it as $H^{\rm pzm}$ and the associated fields as
$\hat{\psi}=(\psi^{0}, \psi^{1})^{\rm t}$. 
We write the Coulomb exchange interaction acting within the PZM sector 
as $V_{\rm X}$ and take $H^{\rm pzm} + V_{\rm X} \equiv H^{\rm eff}$ 
as an effective Hamiltonian that governs the sector for $0\le N_{\rm f} \le 2$.

Let us now suppose filling the PZM levels with electrons gradually 
and examine how level mixing proceeds via the Coulomb interaction.
To this end we rotate $\hat{\psi} =(\psi^{0}, \psi^{1})^{\rm t}$ 
to $\Phi =(\Phi^{0}, \Phi^{1})^{\rm t}$ by an SU(2) matrix, 
\begin{equation}
\hat{\psi} = U \Phi 
=  \left(
\begin{array}{cc}
c_{\theta} & -e^{-i\phi}s_{\theta} \\
e^{i\phi}s_{\theta}&c_{\theta}\\
\end{array}
\right)
\left(
\begin{array}{c}
\Phi^{0}\\ 
\Phi^{1} \\
\end{array}\right),
\label{Psi_UPhi}
\end{equation}
where $c_{\theta}\equiv \cos (\theta/2)$ and $s_{\theta}\equiv \sin (\theta/2)$;
$(\theta, \phi)$ are real angles.
We fix $U$ so that the PZM spectra become diagonal for $\Phi= (\Phi^{0},\Phi^{1})^{\rm t}$ and 
refer to the associated levels as $n=(0_{\theta}, 1_{\theta})$ and their filling fractions 
as $N_{n}=(N_{0}, N_{1})$ (with $0\le N_{n} \le 1)$.
We handle the exchange interaction $V_{\rm X}$ in the Hartree-Fock (HF) approximation 
and cast it in the one-body form 
\begin{eqnarray}
V^{\rm HF}_{\rm X}&=& -   \sum_{\bf p} v_{\bf p}\gamma_{\bf p}^{2}\,
{\cal M}^{mk;a}_{\bf p}\, R^{mk;aa}_{\bf 0}, 
\\
{\cal M}^{mk;a}_{\bf p} &=& g^{mn;a}_{\bf -p} (U{\cal N} U^{\dag})^{nj}\, g^{jk;a}_{\bf p},
\end{eqnarray}
where $m,n,j,k$ run over $(0,1)$; ${\cal N}={\rm diag}(N_{0}, N_{1})$.

We thus write $ H^{\rm eff} = H^{\rm pzm} + V^{\rm HF}_{\rm X}$ as 
\begin{eqnarray}
H^{\rm eff} &=& \int dy_{0}\, \hat{\psi}^{\dag} \hat{H}^{\rm eff} \hat{\psi}
= \int dy_{0}\, \Phi^{\dag} {\cal H} \Phi,
\label{Heff}
\\
\hat{H}^{\rm eff}&=& \left(
\begin{array}{cc}
a&f^{\dag}\\
f&b\\
\end{array}
\right) ,\ \ {\cal H} = U^{\dag}\hat{H}^{\rm eff} U,
\label{hat-H-eff}
\end{eqnarray}
where
\begin{eqnarray}
a  &=&  \hat{\epsilon}_{0} - (N_{1}\,G^{01} + N_{0}\,G^{00})\, c_{\theta}^2 
- (N_{0}\,G^{01} + N_{1}\,G^{00})\, s_{\theta}^2, 
  \nonumber\\
b  &=&  \hat{\epsilon}_{1}- (N_{1}\,G^{11}+ N_{0}\,G^{01})\, c_{\theta}^2 
- (N_{0}\,G^{11}+ N_{1}\,G^{01})\, s_{\theta}^2, 
\nonumber\\
f  &=&  e^{i\phi}\,  (N_{1}-N_{0})\, G^{00;11}\, s_{\theta}\, c_{\theta}
+ {\textstyle c_{1} {e\ell\over{\sqrt{2}}} }\,  (E_{y} + iE_{x}), 
\label{abf}
\end{eqnarray}
with $G^{mn}$, etc., defined in Eq.~(\ref{Gmn}); 
$(\hat{\epsilon}_{0}, \hat{\epsilon}_{1})$ stand for the $N_{\rm f}=0$ spectra. 
Here we have introduced 
coupling to a weak uniform in-plane electric field~\cite{fn_one} ${\bf E}=(E_{x},E_{y})$
to $O({\bf E})$, to detect  an electric dipole moment induced by orbital mixing 
(and for another reason to be clear soon).
Clearly, Eq.~(\ref{abf}) suggests setting $\phi = \arctan (E_{x}/E_{y})$; 
$\phi$ thus controls the direction of ${\bf E}$.  For simplicity, we choose $\phi=0$, 
and specifically use field $E_{y}$ and measure current $j_{x}$.

Let us first take a look at the case of no rotation $\theta \rightarrow 0$ (and $E_{y} \rightarrow 0$).
(i) When $\hat{\epsilon}_{1} > \hat{\epsilon}_{0}$, one first fills the $n=0$ level.
The level  gap $b-a$ increases with  $N_{0}$ and then decreases with $N_{1}$, 
but never closes because $G^{00} > G^{11} > G^{01} >0$ holds. 
Filling the two levels in this way thus realizes a stable configuration. 
(ii) When $\hat{\epsilon}_{0} > \hat{\epsilon}_{1}$, the $n=1$ level is first filled.
The level spectra $(a, b)$ then cross before $N_{\rm f}=2$ is reached. 
This means that a variation in $\theta$ is inevitable to reach the lowest-energy configuration.

Let us therefore suppose  $\hat{\epsilon}_{0} > \hat{\epsilon}_{1}$ and 
try to diagonalize ${\cal H} = U^{\dag} \hat{H}^{\rm eff} U$.
Diagonalization of  ${\cal H}$ is achieved for 
\begin{equation}
{\cal H}^{10} = e^{i\phi}\{ F\, (c_{\theta}^2- s_{\theta}^2) -(a-b)\, s_{\theta} c_{\theta} \}
\rightarrow 0, 
\label{H_zero_one}
\end{equation}
where $F\equiv  G^{00;11}\, (N_{1}-N_{0})\, s_{\theta}\, c_{\theta} + X_{E}$
and $X_{E} \equiv  c_{1} e\ell\,E_{y}/\sqrt{2}$.
In direct calculations it is helpful to isolate 
the difference between $n=\{0,1\}$ modes by setting 
\begin{eqnarray}
\delta \hat{\epsilon}   &=&  \hat{\epsilon}_{0} - \hat{\epsilon}_{1}, \ \ 
 \delta G = G^{00} - G^{11}.
\nonumber\\
\Xi &=&  \delta \hat{\epsilon}  -  (N_{1} c_{\theta}^2 + N_{0}s_{\theta}^2)\, \delta G,
\nonumber\\
D &=& - (G^{01} - G^{00} + G^{00;11}).
\label{XiDelta}
\end{eqnarray}
Equation~(\ref{H_zero_one}), cast in the form 
\begin{equation}
\sin \theta\, \{ \Xi +(N_{1}-N_{0})\,  D\, 
\cos \theta \}  = 2X_{E}\,\cos \theta ,
\label{theta_Nf}
\end{equation}
fixes angle  $\theta$ as a function of filling factor $N_{\rm f} = N_{1}+ N_{0}$.
Minimization [with respect to $(\theta, \phi)$] of the HF ground state energy
also leads to the same equation.

Actually, for bilayer graphene, the combination $D$ vanishes identically, 
$D \rightarrow 0$,
since  $|g^{10}_{\bf p}|^2 - 1 + g^{11}_{\bf p}=0$ holds, as seen from Eq.~(\ref{gmnPZM}).
Here we keep $D$ for a later generalization, 
and refer to the case of bilayer graphene by showing the $D \rightarrow 0$ limit.
For bilayer graphene we set
\begin{equation}
\delta \hat{\epsilon} \rightarrow (1 - \xi)\, \epsilon_{\rm Ls},\   
\delta G \rightarrow 2\, \epsilon_{\rm Ls}\  {\rm and}\ 
D \rightarrow 0. 
\end{equation}

Note first that, for $E_{y}\not= 0$, $\sin\theta=0$ is not a solution to Eq.~(\ref{theta_Nf}). 
This means that, as $N_{\rm f}$ is increased from 0 to 2, 
$\theta$ lies in either domain $0 < \theta <\pi$ or $-\pi<\theta <0$. 
In particular, for $N_{\rm f} \rightarrow 0$, 
Eq.~(\ref{theta_Nf}) yields 
\begin{equation}
\theta \equiv  \delta\theta^{E}_{-} =2 X_{E}/\delta \hat{\epsilon} \ \ ({\rm mod}\ \pi),
\end{equation}
while, for $N_{\rm f} \rightarrow 2$, one finds
\begin{equation}
\theta \equiv  - \delta\theta^{E}_{+} = - 2 X_{E}/(\delta G -\delta \hat{\epsilon}) \ \ ({\rm mod}\ \pi);
\label{dtheta-plus}
\end{equation}
$\delta\theta^{E}_{\mp}
\stackrel{D\rightarrow 0}{\rightarrow} 2 (X_{E}/\epsilon_{\rm Ls})/(1\mp \xi)$.
With $\delta \hat{\epsilon} > 0$ and $|\xi|<1$, 
this implies the following: 
$\theta$ takes the same sign as $E_{y}$, and $|\theta|$ rises from 
$|\delta \theta^{E}_{-}|$ to $\pi - |\delta\theta^{E}_{+}|$ 
as $N_{\rm f}$ is varied from 0 to 2.
One can thus select the sign of $\theta$ 
by turning on a weak Hall field $E_{y}$ or letting a current $j_{x}$ flow.
[Incidentally, for $\hat{\epsilon}_{1} > \hat{\epsilon}_{0}$, 
one finds the solution $\theta = O(X_{E}/\delta \epsilon)$, 
with  the spectra given by $(a,b)|_{\theta\rightarrow 0}$, 
apart from negligible corrections of $O({\bf E}^2)$. ]

Actually, the $\theta >0$  and $\theta <0$ solutions for $U$ are related by a unitary transformation,
$U|_{-\theta} =Y U Y^{-1}$ with $Y= e^{i\pi\, \sigma_{3}/2} = i\sigma_{3}$. 
This fact reflects the invariance of the system under a rotation by angle $\pi$ of coordinates
[or ${\bf x}=(x,y) \rightarrow -{\bf x}]$, as seen from the associated change of 
$\hat{H}^{\rm eff}$  in Eq.~(\ref{Heff}), 
$Y \hat{H}^{\rm eff} Y^{-1} =  \hat{H}^{\rm eff}|_{-f} =  \hat{H}^{\rm eff}|_{-\theta, - {\bf E}}$.
$U$~is thus naturally defined for $-\pi \le \theta \le \pi$.
In view of the anti-periodicity $U|_{\theta + 2\pi} = - U|_{\theta}$ in $\theta$, 
one can even extend $U$ over the full line $-\infty < \theta < \infty$.

For a given angle $\theta$, the level spectra 
$({\cal H}^{00}, {\cal H}^{11}) \equiv  (\hat{\epsilon}_{0_{\theta}}, \hat{\epsilon}_{1_{\theta}})$
are cast in the following two equivalent forms
\begin{eqnarray}
\hat{\epsilon}_{0_{\theta}}
&=&  \hat{\epsilon}_{0}  - N_{1} G^{01} - N_{0} G^{00}
-\Lambda\, s_{\theta}^2 +  (s_{\theta}/c_{\theta})\, X_{E},
\nonumber\\
\hat{\epsilon}_{1_{\theta}}
&=& \hat{\epsilon}_{1} - N_{1} G^{11}  - N_{0} G^{01}
 - \Gamma\, s_{\theta}^2  - (s_{\theta}/c_{\theta})\,X_{E},\ \ 
\label{Spec-s}
\end{eqnarray}
 and 
 \begin{eqnarray}
\hat{\epsilon}_{0_{\theta}}
&=& \textstyle \hat{\epsilon}_{1}- N_{1} G^{01} - N_{0} G^{11}  
+\Gamma\, c_{\theta}^2  +  (c_{\theta}/s_{\theta})\, X_{E} ,
\nonumber\\
\hat{\epsilon}_{1_{\theta}}
&=& \textstyle  \hat{\epsilon}_{0}  - N_{1} G^{00} - N_{0} G^{01}
+\Lambda\, c_{\theta}^2 -  (c_{\theta}/s_{\theta})\, X_{E} ,\ \ 
\label{Spec-c}
\end{eqnarray}
where
\begin{eqnarray}
\Lambda &=& (N_{1}-N_{0})\, D \  \stackrel{D \rightarrow 0}{\rightarrow}\  0,
\nonumber\\
\Gamma &=& (N_{1}-N_{0})(\delta G - D) 
\stackrel{D \rightarrow 0}{\rightarrow}  2\, (N_{1}-N_{0})\,  \epsilon_{\rm Ls}.
\end{eqnarray}
See Appendix A for a derivation of these spectra.

Equation~(\ref{Spec-s}) shows how the spectra 
$(\hat{\epsilon}_{0_{\theta}}, \hat{\epsilon}_{1_{\theta}})$ 
deviate from the $\theta=0$ spectra (of no rotation) 
as $\theta$ grows from zero. 
In contrast, Eq.~(\ref{Spec-c}) shows how they approach, 
as $\theta \rightarrow \pm \pi$ or $c_{\theta} \rightarrow 0$,
the $\theta = \pm \pi$ spectra, 
which, as $N_{\rm f} \rightarrow 2$, attain the filled-level spectra in Eq.~(\ref{Spec_filled}), 
\begin{equation}
(\hat{\epsilon}_{0_{\theta}}, \hat{\epsilon}_{1_{\theta}})|_{\theta \rightarrow \pm\pi}
\rightarrow (\hat{\epsilon}_{1}|^{N_{\rm f}=2}, \hat{\epsilon}_{0}|^{N_{\rm f}=2}).
\label{Spec_pi}
\end{equation}
Thus the empty $n=(0,1)$ levels, when filled, turn into the $n=(1,0)$ levels.

 \begin{figure}[bpt]
\begin{center}
\includegraphics[scale=1.1]{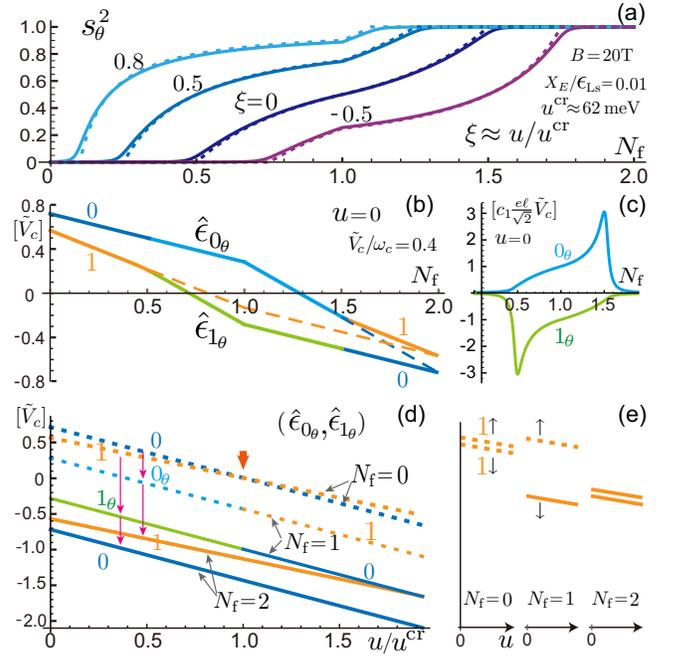}
\end{center}
\vskip-.7cm
\caption{
Orbital mixing in bilayer graphene.  
(a)~Evolution of $s_{\theta}^2= \sin^2(\theta/2)$ with filling $N_{\rm f} =0 \rightarrow 2$ 
at bias $u = (-0.5, 0, 0.5, 0.8)\, u^{\rm cr}$, 
with $X_{E}/\epsilon_{\rm Ls} =0.01$; dotted curves refer to the $E_{y} \rightarrow 0$ limit.
$\theta$ starts to grow around $N_{\rm f} \sim N_{1}^{-}= {1\over{2}}(1- \xi)$.
(b)~Evolution of PZM spectra $(\hat{\epsilon}_{0_{\theta}}, \hat{\epsilon}_{1_{\theta}})$ with filling at $u=0$.
Dashed lines refer to the case of no mixing $\theta=0$. 
(c)~Electric dipole moment per electron induced by orbital mixing. 
(d)~Evolution of PZM spectra with bias $u$ for the empty, half-filled and filled sector $(N_{\rm f}=0,1,2)$.
Dotted lines refer to empty levels. 
In orbital mixing, both levels $\{0_{\theta},1_{\theta}\}$ get shifted with filling $N_{\rm f}$, 
in contrast to the case of spin splitting, depicted in (e).
}
\end{figure}

Figure 2(a) depicts, for bilayer graphene at $B=20\,$T, 
how $(s_{\theta})^2 = \sin^{2}(\theta/2)$ grows as a function of filling factor $N_{\rm f}$ 
for certain values of bias $u$; 
there Eq.~(\ref{theta_Nf}) is  numerically solved for $\theta$, 
with the choice $X_{E}/\epsilon_{\rm Ls} = 0.01$ and $\tilde{V}_{c}/\omega_{c} =0.4$.
$\theta$ starts to rise around $N_{\rm f} \sim {1\over{2}}(1- u/u^{\rm cr})$. 
Figure 2(b) illustrates how PZM spectra $(\hat{\epsilon}_{0_{\theta}}, \hat{\epsilon}_{1_{\theta}})$ 
avoid a crossing via orbital mixing
as they evolve with increasing $N_{\rm f}$.
A sizable gap arises at half-filling $N_{\rm f}=1$, with 
$(\hat{\epsilon}_{0_{\theta}}, \hat{\epsilon}_{1_{\theta}})|^{N_{\rm f}=1} 
= (\hat{\epsilon}_{0}  - G^{01}, \hat{\epsilon}_{0}  -  G^{00})$ for $X_{E}\rightarrow 0$.
Also depicted in Fig.~2(c) is the profile of electric dipole moment induced by orbital mixing,
calculated by numerically differentiating the spectra with respect to $E_{y}$.

Figure 2(d) shows how the PZM spectra evolve 
with bias $u$ for the empty, half-filled and filled sector $(N_{\rm f}=0,1,2)$, respectively.
Dotted lines refer to empty levels. 
In orbital mixing, both levels $\{0_{\theta},1_{\theta}\}$ get shifted with filling $N_{\rm f}=0\rightarrow 2$.
This is in sharp contrast to the case of Coulomb-enhanced spin splitting, depicted in Fig.~2(e).

It is seen from Eqs.~(\ref{Spec-s}) and~(\ref{Spec-c}) 
that the level spectra at angle $\theta$ and at $\pi -\theta$ have reciprocity of the form
\begin{equation}
(\hat{\epsilon}_{0_{\theta}}, \hat{\epsilon}_{1_{\theta}})|_{\theta \rightarrow \pi -\theta}
= (\hat{\epsilon}_{1_{\theta}}, \hat{\epsilon}_{0_{\theta}})|^{\theta}_{N_{0} 
\leftrightarrow N_{1}, -X_{E}}.
\label{Reciprocity}
\end{equation}
This reciprocity derives from the invariance of the basic Hamiltonian 
$\hat{H}^{\rm eff}$ [in Eq.~(\ref{hat-H-eff})],
\begin{equation}
\hat{H}^{\rm eff}|_{\theta \rightarrow \pi -\theta,  N_{0} \leftrightarrow N_{1}, -{\bf E}} = \hat{H}^{\rm eff},
\end{equation}
under simultaneous replacement $\theta \rightarrow \pi -\theta$ (i.e., $s_{\theta} \leftrightarrow c_{\theta}$), 
$N_{0} \leftrightarrow N_{1}$ and $X_{E}  \rightarrow - X_{E}$.

Equation~(\ref{Reciprocity}) implies, in particular, that the $\theta = \pi$ spectra 
 $(\hat{\epsilon}_{0_{\theta}}, \hat{\epsilon}_{1_{\theta}})|_{\theta\rightarrow \pi}$ [in Eq.~(\ref{Spec_pi})]
are equal to the $\theta =0$ spectra 
$(\hat{\epsilon}_{1_{\theta}}, \hat{\epsilon}_{0_{\theta}})|_{\theta\rightarrow 0}$ of Eq.~(\ref{Spec-s})
with $N_{1} \leftrightarrow N_{0}$  and $E_{y}\rightarrow -E_{y}$.
Interchanging $N_{1}$ and $N_{0}$ is to adopt, for a given $N_{\rm f}=N_{1}+N_{0}$, 
the filling sequence of the $u>u^{\rm cr}$ case (of  $\delta \hat{\epsilon} <0$ and no rotation). 
From this follows an important observation: 
The $\theta=\pi$ spectra with $u \rightarrow u^{\rm cr}$ upward are equal to 
the $\theta=0$ spectra with $u \rightarrow u^{\rm cr}$ downward. 
The PZM spectra, when regarded as a function of $N_{\rm f}$ and bias $u$, 
are thus smoothly connected across $u=u^{\rm cr}$; we will see an example later.

To see how $\theta$ depends on filling $N_{\rm f}$ explicitly
(apart from its sign)
one can simply  set $E_{y} \rightarrow 0$ in Eq.~(\ref{theta_Nf}).
One then finds either $\sin \theta=0$ or, if  $\sin \theta\not =0$,
\begin{equation}
 \Xi + (N_{1}-N_{0})\, D\, \cos \theta=0,
\end{equation}
which yields 
\begin{equation}
s_{\theta}^2 = {N_{1} \delta G -\delta \hat{\epsilon} - (N_{1}-N_{0}) D
\over{(N_{1}-N_{0})( \delta G -2 D) }}.
\label{stheta-square}
\end{equation}
Note that 
$s_{\theta}^2=0$ for $N_{1} = N_{1}^{-}$
and 
$s_{\theta}^2=1$ (i.e., $\theta =\pm \pi$) for $N_{0} = N_{0}^{+}$, with
\begin{eqnarray}
 N_{1}^{-} &\equiv& (\delta \hat{\epsilon} - N_{0}D)/ (\delta G- D)
\stackrel{D \rightarrow 0}{\rightarrow} {\textstyle{1\over{2}}} (1-\xi), 
\nonumber\\
 N_{0}^{+} &\equiv& (\delta \hat{\epsilon} - N_{1}D)/ (\delta G- D)
\stackrel{D \rightarrow 0}{\rightarrow} {\textstyle{1\over{2}}} (1-\xi).
\end{eqnarray}
In terms of these Eq.~(\ref{stheta-square}) is neatly expressed as  
\begin{equation}
s_{\theta}^2 
= {N_{1} -N_{1}^{-}\over{N_{1}-N_{0}}}\, f_{D},\ \ 
c_{\theta}^2 = {N_{0}^{+} -N_{0}\over{N_{1}-N_{0}}}\, f_{D},
\label{s-theta_c-theta}
\end{equation}
where $f_{D}\equiv (\delta G- D)/(\delta G- 2D) 
\stackrel{D \rightarrow 0}{\rightarrow} 1$.

With $\delta \hat{\epsilon} > 0$, 
filling of the empty PZM sector starts with the $1_{\theta}$ level.
The angle $\theta$ shows different behavior in the following three domains,
\begin{eqnarray} 
 {\rm (i)}\  &&0\le N_{\rm f}  \le N_{1}^{-}, \ \ [\theta =0; N_{0}=0],
\nonumber\\
{\rm (ii)}\  &&N_{1}^{-} \le N_{\rm f}  \le 1+ N_{0}^{+},\ \ [s_{\theta}^2 = 0 \rightarrow 1], 
\nonumber\\
{\rm (iii)}\   &&1+ N_{0}^{+} < N_{\rm f} \le  2,  \ \  [ |\theta| = \pi; N_{1}=1]. 
\label{Three_Domains}
\end{eqnarray}
$\theta$ remains 0 as the $1_{\theta}$ level is filled over domain (i), 
and rises to $\pi$ (or $-\pi)$ through domain (ii), retaining $s_{\theta}=1$ thereafter over (iii).
As seen from Fig.~2(a), the effect of $X_{E} \propto E_{y}$ is noticeable~\cite{fnone} 
only near the boundaries of domain (ii).

Empty  PZM levels $n=\{0,1\}$
have a crossing when one sweeps bias $u$ across $u^{\rm cr}$. 
It is enlightening to see how they avoid a crossing when one of them is partially filled. 
Let us take $N_{\rm f} = N_{1} <0.5$ and try to solve Eq.~(\ref{theta_Nf}) for $\theta$,
with $X_{E}/\epsilon_{\rm Ls}=0.01$ chosen. 
As seen from Fig.~3, $\theta$ starts to rise  around $u/ u^{\rm cr} \sim 1- 2N_{\rm f}$
 and reaches $\pi$ for $u \gtrsim u^{\rm cr}$, with a rapid rise near $u \sim u^{\rm cr}$ for  $N_{\rm f}\ll 1$. 
The associated spectra $(\hat{\epsilon}_{0_{\theta}}, \hat{\epsilon}_{1_{\theta}})$ always stay apart, 
and the $u\ll u^{\rm cr}$ spectra $(\hat{\epsilon}_{0}, \hat{\epsilon}_{1})|_{N_{1}=N_{\rm f}}$ 
are smoothly connected to
the  $u \gtrsim u^{\rm cr}$ spectra $(\hat{\epsilon}_{1}, \hat{\epsilon}_{0})|_{N_{0}=N_{\rm f}}$ 
across $u=u^{\rm cr}$.
In this way, a level gap generally develops with filling via interaction.  
In practice, however, for small filling $N_{\rm f}\ll 1$ an emerging small gap
will be readily washed away by disorder and finite temperature.
One will then observe a collapse of the  quantized conductance around $u\sim u^{\rm cr}$, 
noticing as if a level crossing had taken place.

 \begin{figure}[pt]
\begin{center}
\includegraphics[scale=0.6]{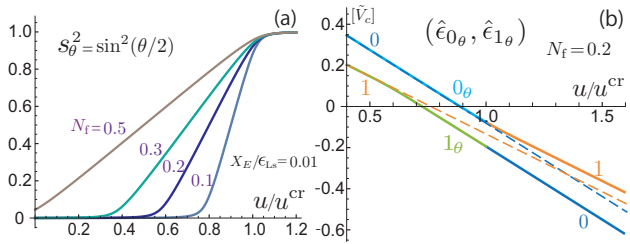}
\end{center}
\vskip-.7cm
\caption{
Evolution of PZM spectra $(\hat{\epsilon}_{0_{\theta}}, \hat{\epsilon}_{1_{\theta}})$ 
with bias $u$ at  small filling $N_{\rm f}<1$. 
(a)~$\theta$ rapidly rises to $\pi$ as $u\rightarrow u^{\rm cr}$ upward.
(b)~The spectra generally avoid a crossing via level mixing even for $N_{\rm f} \ll 1$.
Dashed lines refer to the $\theta=0$ case.}
\end{figure}

We have so far handled mixing of PZM levels in bilayer graphene.
It will be clear now that an inversion of spectra, such as  
$(\hat{\epsilon}_{m} > \hat{\epsilon}_{n})|^{\rm empty}
\rightarrow  (\hat{\epsilon}_{m} < \hat{\epsilon}_{n})|^{\rm filled}$,
is induced by a difference in orbital exchange energy that generally has the property
$G^{mm} > G^{nn}> G^{mn} >0$ for $n>m \ge 0$~\cite{fntwo}. 
The present analysis is equally applicable to such a general case of orbital mixing 
by simply replacing orbital labels $(0,1) \rightarrow (m,n)$ with $m<n$.

\section{non-Abelian Berry's phase}

In this section we wish to clarify algebraic features of orbital-mixing phenomena.
Let us first note that, once ${\cal H} = U^{\dag} \hat{H}^{\rm eff} U$ is diagonalized,
$U$ and ${\cal H}$ are fixed as a function of filling factor $N_{\rm f} = N_{1}+ N_{0}$.
Suppose now that we start filling the empty PZM sector by increasing $N_{\rm f}$
gradually in time, i.e., we set $N_{\rm f} \rightarrow N_{\rm f}(t)$, and ask how the sector evolves.
The eigenmodes of $\hat{H}^{\rm eff}=\hat{H}^{\rm eff}|_{N_{\rm f}\rightarrow N_{\rm f}(t)}$ 
in each instant are thereby written as  
$\hat{\psi}^{(n)}(t) =U|_{N_{\rm f}(t)} \Phi^{(n)} \equiv U(t) \Phi^{(n)}$ 
with $\Phi^{(0)} = (1,0)^{\rm t}$ and $\Phi^{(1)} = (0,1)^{\rm t}$,
and have nondegenerate spectra 
$({\cal H}^{00}, {\cal H}^{11}) = (\hat{\epsilon}_{0_{\theta}}, \hat{\epsilon}_{1_{\theta}})|_{N_{\rm f}(t)}$.

The time evolution of the PZM levels is best clarified 
by referring to the Lagrangian (or action) 
\begin{equation}
L= \int dt dy_{0}\, \hat{\psi}^{\dag}(i\partial_{t} - \hat{H}^{\rm eff}) \hat{\psi}.
\end{equation}
Rewriting $L$ in terms of $\Phi= U^{\dag}\hat{\psi}$ yields
\begin{equation}
L =  \int dt dy_{0}\, \Phi^{\dag} \{ i\partial_{t} - {\cal H} + i (U^{\dag}\partial_{t}U) \} \Phi,
\label{LPhi}
\end{equation}
which describes how the field $\Phi= (\Phi^{0}, \Phi^{1})^{\rm t}$, 
expanded in instantaneous eigenmodes $\{\Phi^{(0)}, \Phi^{(1)}\}$, evolves in time.
It tells us that the field $\Phi$ and associated $(0_{\theta}, 1_{\theta})$ levels
have excitation spectra $ (\hat{\epsilon}_{0_{\theta}}, \hat{\epsilon}_{1_{\theta}})|_{N_{\rm f}(t)}$ 
over the instantaneous ground state 
that evolves along a nontrivial path of mixing [specified by $U(t)$] in the $(\psi^{0}, \psi^{1})$ space.

Here we notice a non-Abelian Berry phase~\cite{WZ}  
 ${\cal A}= -i (U^{\dag}\partial_{t}U) = i(\partial_{t}U^{\dag}) U$, or  the SU(2) connection
 \begin{eqnarray}
 {\cal A}\, dt &=& -i (U^{\dag}\partial_{t}U) dt =  {\cal A}_{\theta} d\theta +  {\cal A}_{\phi} d\phi,
 \\
 {\cal A}_{\kappa} &=&  -i U^{\dag}\partial_{\kappa}U\ \  (\kappa = \theta, \phi). \ \  
 \end{eqnarray}
 In terms of Pauli matrices $\sigma_{a}$, 
 ${\cal A}_{\kappa} =\sum_{a=1}^{3} A^{a}_{\kappa}\sigma_{a}/2$, 
 with   
 \begin{eqnarray}
 A^{a}_{\theta} &=& (\sin \phi, -\cos \phi, 0),
 \nonumber\\
  A^{a}_{\phi} &=& (\cos \phi \sin \theta, \sin \phi \sin \theta, 1- \cos \theta). 
 \end{eqnarray}
Formally the rotation $U^{\dag} = U^{\dag}(t)$ is written 
as a time-ordered (or path-ordered ${\cal P}$) integral of ${\cal A}$, 
\begin{equation}
U^{\dag}(t) = {\cal P}\exp\Big[ -i\! \int^{t}_{0}\! dt\,  {\cal A}\Big]U^{\dag}(0).
\end{equation}
For $\phi=0$, in particular, $ A^{a}_{\theta} \rightarrow (0, -1, 0)$ and
\begin{equation}
U^{\dag} \stackrel{\phi=0}{=} 
e^{i\theta \sigma_{2}/2} = c_{\theta} {\bf 1} + is_{\theta} \sigma_{2}
\end{equation}
is fixed by a net adiabatic change of $\theta = \theta|_{N_{\rm f}(t)}$ alone.
The SU(2) gauge field ${\cal A}_{\kappa}$, associated with filling of the PZM levels, 
derives from interaction $V^{C}$ and resides in the space of parameters $(\theta, \phi)$, 
through which one can adiabatically change the filling factor $N_{\rm f}(t) = N_{1}(t) + N_{0}(t)$, 
interlayer bias $u(t)$, electric field ${\bf E}(t)$, etc.

Let us now recall how $\theta$ behaves in domains (i)-(iii) of Eq.~(\ref{Three_Domains}) 
and reexamine the evolution of the PZM sector (for $\delta \hat{\epsilon}>0$ and $u< u^{\rm cr}$). 
To choose the sign of $\theta$ we start filling the empty sector 
by gradually turning on a weak field $E_{y}>0$ in domain (i), 
and turn it off later before the sector is filled. 
The angle $\theta$ then rises from $0$ to $\pi$ 
as $N_{\rm f} =0 \rightarrow 2$ over time interval $T$, so that    
\begin{equation}
U^{\dag}|_{t=0} = {\bf 1} 
\rightarrow U^{\dag}|_{t=T} = e^{i \pi \sigma_{2}/2} 
= \left(
\begin{array}{cc}
0 & 1 \\
-1& 0 \\
\end{array}
\right) = i \sigma_{2}.
\end{equation}
The instantaneous eigenmodes $\Phi = (\Phi^{0}, \Phi^{1})^{\rm t} =U^{\dag} \hat{\psi}$ 
thereby evolve as mixtures of $(\psi^{0},\psi^{1})$ without a crossing, 
\begin{equation}
\Phi|_{\theta=0}^{\rm empty} =(\psi^0, \psi^{1})^{\rm t}
\rightarrow \Phi|_{\theta=\pi}^{\rm filled} =(\psi^1, - \psi^{0})^{\rm t},
\end{equation}
bringing the empty $n=(0,1)$ levels to filled $n=(1,0)$ levels eventually.

Let us here note that physically the same consequence of orbital mixing is also reached 
by the $\theta<0$ solution 
$U|_{-\theta} =\sigma_{3} U|_{\theta} \sigma_{3}$, with the evolution 
$(\psi^0, \psi^{1})|^{\rm empty} \rightarrow -(\psi^1, -\psi^{0})|^{\rm filled}$.
In view of this arbitrariness in global phases of $(\psi^0, \psi^{1})$, one can adopt
\begin{equation}
{\rm tr}[U] = {\rm tr}[U^{\dag}] =2\, c_{\theta} = 2 \cos (\theta/2)
\end{equation}
as a measure to characterize the presence or absence of orbital mixing. 
For the present $\theta=\pm \pi$ rotation one has ${\rm tr}[U] = 0$
while ${\rm tr}[U] =2$ for no mixing ($\theta = 0$).

 \begin{figure}[bpt]
\begin{center}
\includegraphics[scale=0.95]{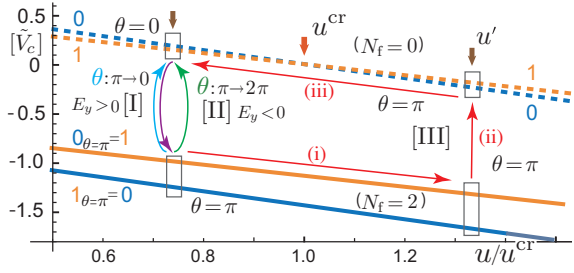}
\end{center}
\vskip-.7cm
\caption{Evolution of PZM spectra $(\hat{\epsilon}_{0_{\theta}}, \hat{\epsilon}_{1_{\theta}})$ 
with an adiabatic change in $(N_{\rm f}, u, E_{y})$ and the associated path-dependence 
of the Berry phase.   }
\end{figure}

We have reached the filled PZM sector. 
Let us next consider returning to the empty sector by further adiabatic change $N_{\rm f} = 2 \rightarrow 0$. 
There are a number of ways to do so, that display the basic character,
the path dependence, of the Berry phase (factor) $U$.   See Fig.~4.

Case [I]: Pass through domain (ii) with weak field $E_{y}>0$ turned on as before.
One then comes back to the original empty sector with a net variation 0 in $\theta$, 
i.e., $U={\bf 1}$, ${\rm tr}[U]=2$ and 
$(\psi^{0}, \psi^{1})|_{t=0}^{\rm empty} \rightarrow (\psi^{0}, \psi^{1})|_{t= T+T'}^{\rm empty}$.

Case [II]:   Turn on a weak field (reversed in sign) $E_{y}< 0$ gradually in coming down through domain (iii). 
Then $\theta$ increases across $\pi$, as seen from Eq.~(\ref{dtheta-plus}), 
and continues to rise as $N_{\rm f}$ is further reduced. 
Turning $E_{y}$ off later in domain (i) takes one to the empty sector 
with a net variation $2\pi$ in $\theta$ and 
\begin{equation}
U^{\dag}|_{\theta = 2\pi}=  e^{2\pi i\sigma_{2}/2} =- {\bf 1}, \ \   {\rm tr}[U]=-2.
\label{Theta_twopi}
\end{equation}
$\hat{\psi}=(\psi^{0}, \psi^{1})^{\rm t}$ thus flips sign, 
$\Phi|_{\theta=0}^{\rm empty} =\hat{\psi} \rightarrow \Phi|_{\theta=2\pi}^{\rm empty} = -\hat{\psi}$,  
simply because $\hat{\psi}$ has made a $2\pi$ rotation relative to $\Phi$ 
in the spinor space.

Case [III]: 
(i)~At $N_{\rm f}=2$ and $\theta =\pi$, 
increase first bias $u$ to a value somewhat above the critical bias,  
$u\rightarrow  u' > u^{\rm cr}$.
(ii)~Then reduce $N_{\rm f}$ to zero gradually.  
Filled levels $\{0_{\theta}, 1_{\theta}\}|_{\theta=\pi}= \{1, 0\}|_{u}$ 
thereby evolve into empty levels $\{1, 0\}|_{u'}$ without a mixing and crossing,  
and the associated spectra 
$(\hat{\epsilon}_{0_{\theta}}, \hat{\epsilon}_{1_{\theta}})|_{\theta=\pi}
= (\hat{\epsilon}_{1}, \hat{\epsilon}_{0})$ change as
$(\hat{\epsilon}_{1}, \hat{\epsilon}_{0})|_{u}^{N_{\rm f}=2}
\stackrel{\rm (i)}{\rightarrow} 
(\hat{\epsilon}_{1}, \hat{\epsilon}_{0})|_{u'}^{N_{\rm f}=2}
\stackrel{\rm (ii)}{\rightarrow} (\hat{\epsilon}_{1}, \hat{\epsilon}_{0})|_{u'}^{N_{\rm f}=0}$.
Note here that, with $E_{y}\rightarrow 0$, $\theta$ does not change with bias $u$ 
for the empty or filled sector ($N_{\rm f}=0,2$)  and also for $u>u^{\rm cr}$.

(iii)~Finally bring bias $u'$ back to the  original value $u < u^{\rm cr}$.
The spectra thereby cross,  
$(\hat{\epsilon}_{1} > \hat{\epsilon}_{0})|^{N_{\rm f}=0}_{u'}
\rightarrow (\hat{\epsilon}_{1} <  \hat{\epsilon}_{0})|^{N_{\rm f}=0}_{u}$,
across $u^{\rm cr}$ while $\theta$ stays at $\pi$.
Here we see a crossing of empty levels and no level mixing.
In this way, via the $N_{\rm f}=0 \rightarrow 2 \rightarrow 0$ cyclic path $\equiv {\cal C}$ 
one returns to the empty sector, 
with a net variation $\pi$ in $\theta$ and 
\begin{equation}
U^{\dag}|^{\cal C}_{\theta=\pi} = e^{i \pi \sigma_{2}/2}= i \sigma_{2}, \ \  {\rm tr}[U] =0.
\end{equation}
The initial and final configurations are physically the same 
although they differ in assignment of $\hat{\psi}$ to $\Phi$,
\begin{equation}
\Phi|_{\theta=0}^{\rm empty} =(\psi^0, \psi^{1})^{\rm t}
\rightarrow \Phi|_{\theta=\pi}^{\rm empty} =(\psi^1, - \psi^{0})^{\rm t}.
\end{equation}
The presence of {\em active} (i.e., interaction-induced) orbital level mixing 
is characterized by $\theta=\pi$ and ${\rm tr}[U] =0$.

\section{trilayers}

Trilayer graphene supports $4 \times 3=12$ PZM levels with a three-fold orbital degeneracy.
As discussed theoretically~\cite{GCP,KA,KM83,YRK,ZTM,ks_ABC} and 
observed experimentally~\cite{BZZL,TWTJ,LVTZ,EVT}, 
the Landau-level spectra and electronic properties of trilayers 
strongly depend on the stacking order, such as $ABA$ and $ABC$ stackings. 
The orbital degeneracy is again lifted by the Coulomb interaction and 
the orbital Lamb shift leads to orbital mixing, as noted earlier~\cite{ks_ABC}.  
In this section we summarize and refine the result 
in the light of the present framework of level mixing.

The $ABC$-stacked trilayer is a chiral generalization~\cite{GCP} of bilayer graphene 
and the zero-energy modes residing primarily in outer layers show a degeneracy 
in orbitals $n=(0,1,2)$ per spin and valley.
The one-body spectra $\{\epsilon_{0}, \epsilon_{1}, \epsilon_{2}\}$ 
deviate from zero energy by a symmetric interlayer bias $u$, with slightly different gradients, 
and the two valleys are related, e.g, as $\epsilon_{n}|^{K'} = \epsilon_{n}|^{K}_{-u}$ per spin.
The orbital Lamb-shift corrections read~\cite{ks_ABC}, e.g., 
\begin{equation}
(\epsilon^{\rm Ls}_{0}, \epsilon^{\rm Ls}_{1}, \epsilon^{\rm Ls}_{2}) 
\stackrel{B=10{\rm T}}{\approx} (0.888, 0.777, 0.641) \,  \tilde{V}_{c} 
\label{Evn} 
\end{equation}
numerically,
with only the leading band parameter 
$g=\gamma_{1}/\omega_{c} \approx 3.41$ 
kept at $B=10\,$T.
The PZM spectra $\hat{\epsilon}_{n} =\epsilon_{n} + \epsilon_{n}^{\rm Ls}$
are ordered as $\hat{\epsilon}_{0} > \hat{\epsilon}_{1} >\hat{\epsilon}_{2}>0$ for empty levels
at zero bias $u=0$ while they change sign for filled levels so that 
$0 > \hat{\epsilon}_{2} > \hat{\epsilon}_{1} > \hat{\epsilon}_{0}$ at $N_{\rm f}=3$.
This signals the presence of orbital mixing upon level filling.

To study level mixing let us rotate, as in the bilayer case, the PZM sector
$\hat{\psi} = (\psi^{0}, \psi^{1}, \psi^{2})^{\rm t}$ to $\Phi = (\Phi^{0}, \Phi^{1}, \Phi^{2})^{\rm t}$ 
by an SO(3) matrix $U$, $\hat{\psi} = U \Phi$, 
and try to diagonalize the HF effective Hamiltonian $\hat{H}^{\rm eff}$ 
(taken to be a real symmetric matrix).
Previously we wrote $U$ as a product of three rotations and 
fixed it numerically as a function of filling factor $N_{\rm f}$ at zero bias $u=0$. 
It is illuminating to cast the result in the polar form
\begin{equation}
U = e^{i \chi_{a}\, t_{a}} = e^{i \Theta\, {\bf n\cdot t}}, 
\end{equation}
where ${\bf t}=(t_{0}, t_{1}, t_{2})$ stand for the spin-1 generators $(t_{a})^{bc} \equiv i \epsilon^{bac}$ 
with the totally antisymmetric tensor 
$\epsilon^{abc}$ and $ \epsilon^{012} =1$; 
real angles $\vec{\chi}= (\chi_{0}, \chi_{1}, \chi_{2}) = \Theta\, {\bf n}$ 
are decomposed into the magnitude $\Theta \equiv ~|\vec{\chi}|$ 
and a unit vector  ${\bf n} =(n_{0}, n_{1}, n_{2})$; 
${\bf n\cdot t} = n_{a}t_{a}$.
Note that $\chi_{0}$ mixes $n=(1,2)$, $\chi_{1}$ mixes $(0,2)$, etc.
Some useful formulas are
 \begin{eqnarray}
e^{i \Theta\,  {\bf n\cdot t}} 
&=& 1 + (\cos \Theta -1)\, P + i ({\bf n\! \cdot\! t})\, \sin \Theta, 
\nonumber\\
{\rm tr}[e^{i \Theta\, {\bf n\cdot t}}] &=& 1 +  2\cos \Theta, 
\end{eqnarray}
where  $P \equiv ({\bf n\! \cdot\! t})^{2}$ is a projection operator,
 $P^2=P$ and $({\bf n\! \cdot\! t})P=({\bf n\! \cdot\! t})$; $P^{ab} = \delta_{ab} - n_{a}n_{b}$.

 \begin{figure}[tbp]
\begin{center}
\includegraphics[scale=0.7]{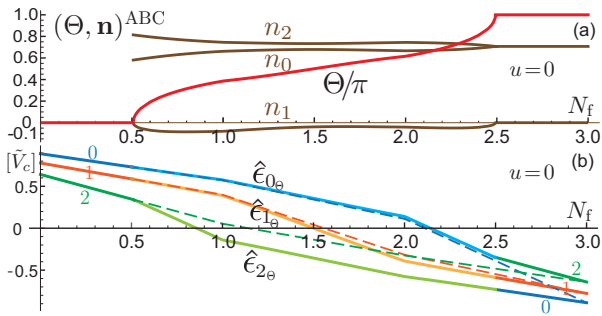}
\end{center}
\vskip-.7cm
\caption{
Orbital mixing in $ABC$ trilayer graphene at $B=10\,$T and $u=0$. 
(a)~Rotation $U=e^{i\Theta\, {\bf n\cdot t}}$.  
(b)~Evolution of PZM spectra 
$(\hat{\epsilon}_{0_{\Theta}}, \hat{\epsilon}_{1_{\Theta}},  \hat{\epsilon}_{2_{\Theta}})$
with filling $N_{\rm f}=0\rightarrow 3$.
Dashed curves refer to the spectra of no mixing $\Theta=0$. 
}
\end{figure}

Figure 5 illustrates how the angle $\Theta$ and direction ${\bf n}$ change with filling factor $N_{\rm f}$.
$\Theta$ rises from zero  to $\pi$ over the interval $0.51 \lesssim N_{\rm f} \lesssim 2.51$
and ${\bf n}$ lies around ${\bf n}|_{\Theta = \pi} = (1,0,1)/\sqrt{2}$. 
It is essential that the three levels cooperate,
with the associated SO(3) Berry phase
${\cal A}_{\Theta} \approx  {\bf n\cdot t}$.
At $N_{\rm f}=3$ and $\Theta=\pi$,
\begin{equation}
U = e^{i \pi {\bf t}\cdot {\bf n}|_{\Theta=\pi}}
= \left(
\begin{array}{ccc}
& &1 \\
&-1&\\
1& & \\
\end{array}
\right)\ {\rm and}\ {\rm tr}(U) = -1. 
\end{equation}
Upon filling, the eigenmodes
$(\Phi^{0}, \Phi^{1}, \Phi^{2})$ thus evolve from
$(\psi^{0}, \psi^{1}, \psi^{2})|^{\rm empty}_{N_{\rm f}=0}$ 
to  $(\psi^{2}, -\psi^{1}, \psi^{0})|^{\rm filled}_{N_{\rm f}=3}$
without a crossing, as seen from Fig.~5(b).

There is another solution 
that differs from one shown in the figure by signs, 
$(n_{0}, n_{1}, n_{2}) \rightarrow (-n_{0}, n_{1}, -n_{2})$.
It is related to $U$ by a unitary transformation, 
$U|_{-n_{0}, n_{1}, -n_{2}} ={\cal Y}\,  U {\cal Y}^{-1}$
with
${\cal Y}= e^{i \pi t_{1}}={\rm diag}[-1,1,-1]$, and
reflects again the invariance of the system under a spatial $\pi$ rotation,
${\bf x} \rightarrow -{\bf x}$.
It is enlightening to interpret $U|_{-n_{0}, n_{1}, -n_{2}}$ 
as a rotation by negative angle $\Theta < 0$ about the axis ${\bf n}' = (n_{0},-n_{1}, n_{2})$.
Then $U = e^{i \Theta\,  {\bf n\cdot t}}$, as a function of $N_{\rm f}$, 
is naturally defined for $- \pi \le \Theta \le \pi$, and 
even for the full line $- \infty < \Theta < \infty$ 
if one notes that $U$ has period $2\pi$ in $\Theta$.
One can also control the sign of $\Theta$ by use of a weak in-plane field $E_{y}$~\cite{ks_ABC}.
It is clear now that $U$ acts as a path-dependent non-Abelian phase factor 
when one controls $(\Theta, {\bf n})$ 
via adiabatic changes of external parameters $(N_{\rm f}, {\bf E}, u, \cdots)$, 
as in Fig.~4 of the bilayer case.

Previously $U|_{\Theta=\pi}$ was obtained as a product of three  $\pi/2$ rotations,
$U = e^{-i (\pi/2) t_{2}}e^{-i (\pi/2) t_{1}} e^{-i (\pi/2) t_{0}}$. 
A single $\Theta=\pi/2$ rotation, e.g,  $e^{ i(\pi/2) t_{0}} = 1 \oplus i\sigma_{2}$, 
consists of a $\theta=\pi$ rotation of $(\psi^{1}, \psi^{2})$, with $\psi^{0}$ left intact.
Such a $|\Theta|=\pi/2$ rotation has been encountered 
in a study~\cite{ks_ABC} of $ABA$-stacked trilayer graphene.
The $ABA$ trilayer accommodates~\cite{GCP,KA} 
monolayer-like and bilayer-like subbands, and 
has the PZM levels specified by orbital labels such as $n =(0,1_{\pm})$ in one valley 
and $n = (0_{\pm},1)$ in another valley.  
It turns out that Coulomb exchange interactions mainly act between $(0,1_{-})$, 
leaving $1_{+}$ rather isolated; analogously for  $(1,0_{-})$ and $0_{+}$.
This explains why $\Theta=\pi/2$ rotations are responsible for mixing of PZM levels in the $ABA$ trilayer.
In this way,  indices ${\rm tr}[U|_{\Theta=\pi}] =-1$ and  ${\rm tr}[U|_{\Theta=\pi/2}] =1$ 
clearly distinguish $ABC$ and $ABA$ trilayers in their character of orbital mixing.

\section{ Evolution and crossing of many-body states}

Bilayer graphene has four {\it renormalized} PZM sectors 
of valley $(K,K')$ $\times$ spin $(\uparrow, \downarrow)$.
Each sector, when empty or filled, becomes an eigenstate to $O(V^{C})$ of the total Hamiltonian 
$H^{\rm 1b}+ V^{C}$, as we have noted. 
Such empty and filled sectors do not mix by exchange interaction to  $O(V^{C})$,
unless there is a degeneracy (i.e., unless they cross). 
In the light of this picture, we discuss in this section 
what the $\nu=0$ ground state is like 
when orbital splitting $\epsilon_{\rm Ls}$ and spin splitting (with Zeeman energy 
$\mu_{\rm Z} \equiv g^{*} \mu_{e}B \sim 0.12\, B[{\rm  T}]$) are taken into account. 
We start with the four empty PZM sectors that constitute the unique ground state 
at total filling factor $\nu=-4$, and consider filling them with electrons gradually.

Whenever nonzero bias $u > 0$ induces a sizable valley gap $\sim u$, 
the $\nu=0$ ground state is certainly realized as a valley-polarized one 
with $(1_{\downarrow}^{K}, 0_{\downarrow}^{K})$ + $(1_{\uparrow}^{K}, 0_{\uparrow}^{K})$ filled 
(in obvious notation) and of total energy (per electron)
\begin{equation}
\epsilon^{\rm (v)}=2\, (\hat{\epsilon}_{0} + \hat{\epsilon}_{1})|^{N_{\rm f}=2},
\end{equation} 
which gets lower with increasing bias $u$.

 \begin{figure}[tbp]
\begin{center}
\includegraphics[scale=0.6]{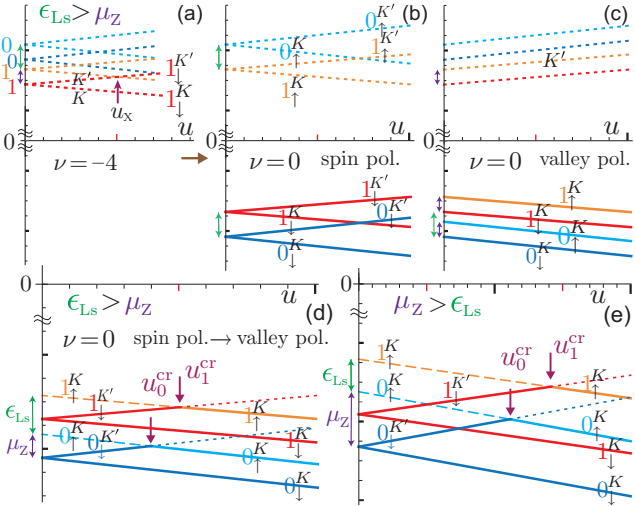}
\end{center}
\vskip-.7cm
\caption{
The $\nu=0$ ground state in bilayer graphene. 
The empty LLL at small bias $u\sim 0$ 
[in (a)], upon half filling, grows into a spin-polarized $\nu=0$ state [in (b)] 
while, at higher bias $u > u_{\rm X}$, it evolves into a valley-polarized state [in (c)]. 
(d)~and (e)~The spin-polarized $\nu=0$ ground state, with increasing bias $u$, 
evolves into the valley-polarized one in two steps at $(u^{\rm cr}_{0}, u^{\rm cr}_{1})$.
}
\end{figure}

For small bias $u \sim 0$, spin splitting will become important.
See Fig.~6.
Figure~6(a) depicts the empty level spectra of  the $\nu=-4$ ground state
for $\epsilon_{\rm Ls}  \gtrsim \mu_{\rm Z}$, 
as is normally the case.
At $u\sim 0$, $(1^{K}_{\downarrow}, 1^{K'}_{\downarrow})$ are lower than others
(for both $\epsilon_{\rm Ls}  \gtrsim \mu_{\rm Z}$ and $\epsilon_{\rm Ls} < \mu_{\rm Z}$).
Accordingly, for  $u \sim 0$, 
the $\nu=-4$ state [in~6(a)], upon filling  $(1_{\downarrow}^{K}, 0_{\downarrow}^{K})$ 
and $(1_{\downarrow}^{K'}, 0_{\downarrow}^{K'})$ in  sequence or in some other order, 
will evolve into a spin-polarized $\nu=0$ ground state [in 6(b)] 
of total energy
\begin{equation}
\epsilon^{\rm (s)} = (\hat{\epsilon}_{0}+ \hat{\epsilon}_{1})|^{N_{\rm f}=2} 
+(\hat{\epsilon}_{0}+ \hat{\epsilon}_{1})|^{N_{\rm f}=2}_{-u} - 2 \mu_{\rm Z},
\end{equation}
which barely depends on  $u$.

Of these two $\nu=0$ candidates 
the spin-polarized state is generally favored for $u \sim 0$, as seen from  
\begin{equation}
\epsilon^{\rm (v)}-\epsilon^{\rm (s)} 
= 2\, (\mu_{\rm Z} - \kappa [u] ) \approx  2\, (\mu_{\rm Z} - \lambda\, u ),
\end{equation}
where $\kappa [u] = (1+z_{1})\,  u/2  - \Omega_{0}^{K} - \Omega_{1}^{K} \approx \lambda\,  u$
and $\lambda = O(1)$.  
The $\nu=0$ ground state, if formed as a spin-polarized one [in 6(b)] at $u\sim 0$, 
will eventually evolve into a valley-polarized state [in 6(c)] as $u$ is increased. 
Let us consider how this transition takes place.
Figure~6(d) depicts the filled spectra of the spin-polarized state with those of the other superposed.
Figure~6(e) shows similar spectra for the case $\mu_{\rm Z} > \epsilon_{\rm Ls}$ in high field $B$. 
Let us first take a look at the latter. 
There, as $u$ is increased, the valley-polarized virtual state comes down in energy 
and, at first,  filled $1^{K'}_{\downarrow}$ meets $0^{K}_{\uparrow}$. 
At this point of degeneracy, $1^{K'}_{\downarrow}$ has a chance to turn into filled $0^{K}_{\uparrow}$,
but this is not possible because filled $1^{K'}_{\downarrow}$ has to first mix 
with empty $0^{K}_{\uparrow}$ which lies far above in the spectra.

Next, filled $0^{K'}_{\downarrow}$ meets $0^{K}_{\uparrow}$ in the figure.  
At this degeneracy, filled $0^{K'}_{\downarrow}$ can readily turn into filled $0^{K}_{\uparrow}$ 
via a global rotation in the valley$\times$spin space.
Note that $g^{00;K'}_{\bf p} = g^{00;K}_{\bf p}=1$ 
so that, within the $n=0$ sector, the Coulomb interaction is invariant under rotations 
in valleys and spins. 
Thus there is no extra cost of energy in making a rotation 
$(0^{K}_{\uparrow}|^{\rm empty}, 0^{K'}_{\downarrow}|^{\rm filled})
\stackrel{U}{\rightarrow}
 (0^{K'}_{\downarrow}|^{\rm empty}, 0^{K}_{\uparrow}|^{\rm filled})$, 
with a non-Abelian Berry phase $U$ of $\theta=\pi$ and ${\rm tr}[U]=0$.
Such a rotation takes place across the critical bias 
\begin{equation}
u^{\rm cr}_{0} = \mu_{\rm Z}/\lambda_{0}, 
\end{equation}
with  $\lambda_{0}= 1-  2\, \Omega_{0}/u \stackrel{B = 20 {\rm T}}{\sim}  1+ 0.65\,  \tilde{V}_{c}/\omega_{c}$.
See Appendix B for details. 
Similarly, filled $1^{K'}_{\downarrow}$ turns into filled $1^{K}_{\uparrow}$ 
with little cost of energy across the second critical bias
\begin{equation}
u^{\rm cr}_{1} = \mu_{\rm Z}/\lambda_{1} >  u^{\rm cr}_{0},
\label{ucr_one}
\end{equation}
with  
$\lambda_{1} = z_{1} -  2\, \Omega_{1}/u  
\stackrel{B = 20 {\rm T}}{\sim}  z_{1}+ 0.61\,  \tilde{V}_{c}/\omega_{c}< \lambda_{0}$.
[Actually, $(\lambda_{0}+ \lambda_{1})/2 = \lambda$, 
and $u^{\rm cr}_{1}$ coincides with $u_{\rm X}$ in Fig.~6(a).] 
The valley-polarized $\nu=0$ ground state is thus reached in the following sequence 
\begin{equation}
(1_{\downarrow}^{K}, 0_{\downarrow}^{K})\oplus \{(1_{\downarrow}^{K'}, 0_{\downarrow}^{K'}) 
\stackrel{u^{\rm cr}_{0}}{\rightarrow}    (1_{\downarrow}^{K'}, 0_{\uparrow}^{K})
\stackrel{u^{\rm cr}_{1}}{\rightarrow}   (1_{\uparrow}^{K}, 0_{\uparrow}^{K}) \}
\label{seq_transition}
\end{equation}
and continues for $u>u^{\rm cr}_{1}$.
For $u \in (u^{\rm cr}_{0},  u^{\rm cr}_{1})$ 
the ground state is polarized in both valley and spin;
this intermediate state differs in structure from one discussed earlier in Refs.~\cite{LC,KJ}.
It is clear from Fig.~6(d) that the transition follows the same steps 
for the case $\epsilon_{\rm Ls}  \gtrsim \mu_{\rm Z}$ as well.
Inclusion of weak $e$-$h$ breaking $(\Delta, \gamma_{4}, \dots)$ also leaves 
this two-step picture qualitatively intact. 

An observable signature of such transitions is the following: With increasing bias $u$,
the quantum Hall effect will survive as long as the ground state and competing virtual state 
retain an appreciable energy gap. 
Incompressibility will be lost and conductance $\sigma_{xx}$ will rise from zero 
only when bias $u$ lies around these critical values $u^{\rm cr}_{0}$ and $u^{\rm cr}_{1}$.
It is clear, on interchanging valleys $K \leftrightarrow K'$, 
that negative bias $u<0$ also leads to the same sequence of transition.

Actually, early transport experiments~\cite{WAFM,KLT} 
observed a collapse of the $\nu=0$ quantum Hall state 
at two distinct (positive/negative) values of electric field $\propto u$, 
and later capacitance measurements~\cite{LFX, HLZW} in higher magnetic field $B$ 
detected it at four such values of $u$. 
The transition sequence in Eq.~(\ref{seq_transition}) appears consistent with one 
inferred from layer-sensitive capacitance measurements of Hunt {\it et al.}~\cite{HLZW}.

\section {summary and discussion}

Characteristic to  few-layer Dirac electron systems in a magnetic field 
is a multiplet, at the LLL, of PZM levels nearly degenerate in orbitals, valleys and spins.  
Their spectra are sensitive to interactions and external perturbations,
and, in particular, the orbital Lamb shift inevitably induces a level inversion 
between the empty and filled levels in a way governed by $e$-$h$ symmetry.

In the present paper we have examined how those PZM levels evolve 
with increasing filling and external bias $u$
under many-body interactions, 
and have seen that they generally avoid a crossing via level mixing 
which is governed by a non-Abelian Berry's phase (factor) $U$. 
This Berry's phase derives from interactions, and encodes, in the form of trace ${\rm tr}(U)$, how a nearly degenerate system responds to adiabatic external changes, 
such as the filling factor, electric and magnetic fields.
Its path dependence, in particular, reveals 
algebraic features underlying general Landau-level crossing/mixing phenomena.  
Our basic picture of level mixing is also applicable to evolution of many-body ground states 
with sweeping external perturbations, 
as examined in Sec.~VI for the $\nu=0$ ground state in bilayer graphene.

Landau-level crossing/mixing phenomena deserve serious attention as a platform to explore, 
both theoretically and experimentally, many-body physics. 
Our focus has so far been on mixing of PZM levels themselves. 
Crossings of PZM levels with other higher levels, 
as observed in ABA trilayer graphene~\cite{TWTJ}, 
deserve equal attention, 
we remark, although a close look into their many-body features is left here for future study.

\acknowledgments
 This work was supported in part by a Grant-in-Aid for
Scientific Research from the Ministry of Education, Science,
Sports and Culture of Japan (Grant No. 21K03534).

\appendix

\section{Rotated level spectra}

In this appendix we outline the derivation of the level spectra in Eqs.~(\ref{Spec-s}) and~(\ref{Spec-c}). 
The diagonal elements of ${\cal H} = U^{\dag}\hat{H}^{\rm eff} U$ 
in Eq.~(\ref{hat-H-eff}) are written as
\begin{eqnarray}
{\cal H}^{00} &=& a\, c_{\theta}^{2}  + b\, s_{\theta}^{2} + 2F s_{\theta} c_{\theta},
\nonumber\\ 
{\cal H}^{11} &=& a\, s_{\theta}^{2}  + b\,  c_{\theta}^{2} - 2F s_{\theta} c_{\theta}, 
\end{eqnarray}
with $F\equiv  (N_{1}-N_{0})\, G^{00;11} s_{\theta}\, c_{\theta} + X_{E}$ and 
$X_{E} = c_{1}e\ell E_{y}/\sqrt{2}$.
Note parametrization in Eq.~(\ref{XiDelta}) for direct calculations.
The spectra are thereby rewritten as 
\begin{eqnarray}
{\cal H}^{00}  &=& \textstyle
 \hat{\epsilon}_{0}  - N_{1} G^{01} - N_{0} G^{00} - X^{0},
 \nonumber\\
{\cal H}^{11}  &=& \hat{\epsilon}_{0}  - N_{1} G^{00} - N_{0} G^{01} - X^{1},
\label{SpecCalH}
\end{eqnarray}
with 
\begin{eqnarray}
X^{0} &=&  \Xi\, s_{\theta}^2 + 2s_{\theta}c_{\theta} (s_{\theta}c_{\theta} \Lambda -  X_{E}),
\nonumber\\
 X^{1} &=&  \Xi\, c_{\theta}^2 -  2s_{\theta}c_{\theta} (s_{\theta}c_{\theta} \Lambda -  X_{E}),  
 \label{R_zero_one}
\end{eqnarray}
where $\Lambda = (N_{1}-N_{0})\, D$.

On the other hand,  
${\cal H}^{10}=0$ [Eq.~(\ref{H_zero_one})] implies
the relation $\Xi = (c_{\theta} ^2-s_{\theta} ^2)(-\Lambda + X_{E}/s_{\theta}c_{\theta} )$.
Substituting this into Eq~(\ref{R_zero_one}) yields
\begin{eqnarray}
X^{0} &=& \Lambda\, s_{\theta}^2 - (s_{\theta}/c_{\theta}) X_{E},
\nonumber\\
X^{1} &=& - \Lambda\, c_{\theta}^2 +  (c_{\theta}/s_{\theta})X_{E}.  
\end{eqnarray}
Note that $X^{0}+ X^{1}= \Xi$, which then reads
\begin{equation}
\delta \hat{\epsilon} \equiv \hat{\epsilon}_{0} -\hat{\epsilon}_{1} 
= (N_{1} c_{\theta}^2 +N_{0} s_{\theta}^2 ) \delta G +  X^{0} + X^{1}.
\label{deltaEps}
\end{equation}
On replacing $\hat{\epsilon}_{0} \rightarrow \hat{\epsilon}_{1}+ \delta \hat{\epsilon}$ 
in Eq.~(\ref{SpecCalH}), 
the spectra are cast in two equivalent forms in Eqs.~(\ref{Spec-s}) and (\ref{Spec-c}).
\\

\section{Global mixing among PZM levels}

In this appendix we examine how global rotations 
$0^{K'}_{\downarrow} \rightarrow 0^{K}_{\uparrow}$ 
and  $1^{K'}_{\downarrow} \rightarrow 1^{K}_{\uparrow}$, posed in Sec.~VI, 
proceed via exchange interaction.
Let us start with the $n=0$ orbital modes, and try to rotate 
a pair of (empty, filled) fields $(\psi^{0;K}_{\uparrow}, \psi^{0;K'}_{\downarrow})^{\rm t}$ 
to $(\Phi^{0}_{\rm e}, \Phi^{0}_{\rm f})^{\rm t}$ 
by a unitary matrix $U(\theta)$, as in Eq.~(\ref{Psi_UPhi}).
As verified readily, the associated HF interaction $V^{\rm HF}_{\rm X}$
takes a simple form 
\begin{equation}
V^{\rm HF}_{\rm X} = - G^{00} ( N^{\rm f} \,{\cal R}_{{\rm ff};{\bf 0}} 
+ N^{\rm e} \,{\cal R}_{{\rm ee};{\bf 0}} )
\end{equation}
in terms of $(\Phi^{0}_{\rm e}, \Phi^{0}_{\rm f})$ 
with filling fractions $(N^{\rm e},N^{\rm f})$ and charge operators 
${\cal R}_{{\rm ff};{\bf 0}}= \int dy_{0} \Phi^{0 \dag}_{\rm f} \Phi^{0}_{\rm f}$, etc.
Setting $(N^{\rm e},N^{\rm f}) \rightarrow (0,1)$ shows that 
global valley$\times$spin rotations of the filled $n=0$ level $(\sim \Phi^{0}_{\rm f})$ 
require no extra cost of Coulombic energy.

The one-body terms with spectra 
$\hat{\epsilon}_{0\uparrow}^{K} = \hat{\epsilon}_{0}^{K} + \mu_{\rm Z}/2$ and 
$\hat{\epsilon}_{0\downarrow}^{K'} = \hat{\epsilon}_{0}^{K'} -  \mu_{\rm Z}/2$ 
are combined with $V^{\rm HF}_{\rm X}$ 
to yield the effective Hamiltonian for the rotated field,
\begin{eqnarray}
H^{\rm eff} &=& 
\{ \epsilon_{\rm f}(\theta) - G^{00}\}\, {\cal R}_{{\rm ff};{\bf 0}} 
+ \epsilon_{\rm e}(\theta)\, {\cal R}_{{\rm ee};{\bf 0}} 
\nonumber\\
&& -s_{\theta}c_{\theta} (\hat{\epsilon}_{0\uparrow}^{K} -\hat{\epsilon}_{0\downarrow}^{K'})  
({\cal R}_{{\rm ef};{\bf 0}} + {\cal R}_{{\rm fe};{\bf 0}} ),
\label{Heff_global}
\end{eqnarray}
where $\epsilon_{\rm f}(\theta)= s_{\theta}^2\, \hat{\epsilon}_{0\uparrow}^{K} 
+ c_{\theta}^2\, \hat{\epsilon}_{0\downarrow}^{K'}$
and $\epsilon_{\rm e}(\theta)= c_{\theta}^2\, \hat{\epsilon}_{0\uparrow}^{K} 
+ s_{\theta}^2\, \hat{\epsilon}_{0\downarrow}^{K'}$.
Diagonalization is therefore achieved for $\theta\equiv 0$ (mod $\pi)$ 
or, if $s_{\theta}c_{\theta}\not=0$, for 
\begin{equation}
\hat{\epsilon}_{0\uparrow}^{K} -\hat{\epsilon}_{0\downarrow}^{K'}
=\mu_{\rm Z} -u + \Omega_{0}- \Omega_{0}|_{-u} 
= \mu_{\rm Z} - \lambda_{0} u\rightarrow 0,
\end{equation}
with 
$\lambda_{0}= 1-  2\, \Omega_{0}/u$.
It is now clear that filled $0^{K'}_{\downarrow}$ turns into filled $0^{K}_{\uparrow}$ 
across $u \sim  u^{\rm cr}_{0}= \mu_{\rm Z}/\lambda_{0}$ 
via a global rotation of angle $\theta=\pi$.

Let us next note that $g^{11;K}_{\bf p} =1- {1\over{2}} c_{1}^{2} \ell^{2}{\bf p}^{2}$ 
and $g^{11;K'}_{\bf p} = g^{11;K}_{\bf p}|_{-u}$ 
barely differ for small $u \sim O(\mu_{\rm Z})$.
The Coulomb interaction, acting within the $n=1$ sector, 
thus remains almost invariant under global valley and spin rotations, 
and $H^{\rm eff}$ in Eq.~(\ref{Heff_global}) 
applies to the transition 
${\rm filled}\ 1^{K'}_{\downarrow} \rightarrow {\rm filled}\ 1^{K}_{\uparrow}$ as well, 
with obvious replacement $G^{00} \rightarrow G^{11}$, 
$\hat{\epsilon}_{0\downarrow}^{K'} \rightarrow \hat{\epsilon}_{1\downarrow}^{K'}$, etc.
The result is summarized in Eq.~(\ref{ucr_one}).




\begin{thebibliography}{99}   

\bibitem{NMMKF} K.~S.~Novoselov, E.~McCann, S.~V.~Morozov, V.~I.~Fal'ko, M.~I.~Katsnelson,
U. Zeitler, D. Jiang, F. Schedin, and A.~K. Geim, Nat. Phys. {\bf 2}, 177 (2006).

\bibitem{OBSHR} T. Ohta, A. Bostwick, T. Seyller, K. Horn, and E. Rotenberg,
Science {\bf 313}, 951 (2006).

\bibitem{MF} E. McCann and V. I. Fal'ko, Phys. Rev. Lett. {\bf 96}, 086805 (2006). 

\bibitem{GCP}  F. Guinea, A. H.  Castro Neto, and N. M. R. Peres, 
Phys. Rev. B {\bf 73}, 245426 (2006).

\bibitem{KA} 
M. Koshino and T. Ando, Phys. Rev. B {\bf 76}, 085425 (2007).


\bibitem{BCNM} 
Y.~Barlas, R.~C\^ot\'e, K.~Nomura, and A.~H.~ MacDonald, 
Phys. Rev. Lett. {\bf 101}, 097601 (2008).

\bibitem{KSpzm}  
K. Shizuya, Phys. Rev. B {\bf 79}, 165402 (2009).   

\bibitem{BCLM} 
Y.~Barlas, R.~C\^ot\'e, J.~Lambert, and A.~H.~ MacDonald, 
Phys. Rev. Lett. {\bf 104}, 096802 (2010).

\bibitem{CLBM} R.~C\^ot\'e, J. Lambert,  Y. Barlas, and A. H. MacDonald,
Phys. Rev. B {\bf 82}, 035445 (2010).

\bibitem{CLPBM} R.~C\^ot\'e, W.~Luo, B.~Petrov, Y. Barlas, and A. H. MacDonald,
Phys. Rev. B {\bf 82}, 245307 (2010).

\bibitem{NL} R. Nandkishore and L. Levitov, 
Phys. Rev. B {\bf 82}, 115124 (2010).

\bibitem{GGJ} E.~V. Gorbar, V.~P. Gusynin, J. Jia, and V.~A. Miransky, 
Phys. Rev. B {\bf 84}, 235449 (2011).

\bibitem{KS_Ls}  K. Shizuya, Phys. Rev. B {\bf 86}, 045431 (2012).

\bibitem{Khari} M.~Kharitonov, Phys. Rev. Lett. {\bf 109}, 046803 (2012).

\bibitem{LC}   J. Lambert and R.~C\^ot\'e, Phys. Rev. B {\bf 87}, 115415 (2013).  

\bibitem{KJ}    A.~Knothe and T.~Jolicoeur, Phys. Rev. B {\bf 94}, 235149 (2016).

\bibitem{KScrBG}  K. Shizuya, Phys. Rev. B {\bf 101}, 195429 (2020).


\bibitem{FMY}
B. E. Feldman, J. Martin, and A. Yacoby, Nat. Phys. {\bf 5}, 889 (2009). 

\bibitem{ZCZJ} Y.~Zhao, P.~Cadden-Zimansky, Z.~Jiang, and P.~Kim, Phys. Rev. Lett. {\bf 104}, 066801 (2010).

\bibitem{WAFM}
R. T. Weitz, M. T. Allen, B. E. Feldman, J. Martin, and A. Yacoby,
Science {\bf 330},  812 (2010). 

\bibitem{MFW}  J. Martin, B. E. Feldman, R. T. Weitz, M. T. Allen, and A. Yacoby, 
Phys. Rev. Lett. {\bf 105}, 256806 (2010).

\bibitem{KLT} 
S.~Kim, K. Lee, and E. Tutuc, 
Phys. Rev. Lett. {\bf 107}, 016803 (2011). 

\bibitem{VJB} J. Velasco Jr, L.~Jing, W.~Bao, Y. Lee, P. Kratz, V. Aji, M. Bockrath, 
C. N. Lau, C. Varma, R. Stillwell, D. Smirnov, Fan Zhang, J. Jung, and A. H. MacDonald,
Nature Nanotech. {\bf 7}, 156 (2012).

\bibitem{MDY} 
P. Maher, C.~R. Dean, A.~F.~Young, T.~Taniguchi, K. Watanabe, K.~L. Shepard, J. Hone,
and P. Kim, Nat. Phys. {\bf 9}, 154 (2013). 

\bibitem{LFX} K. Lee, B.~Fallahazad, J. Xue, D.~C.~Dillen, K.~Kim,
T.~Taniguchi, K.~Watanabe, and E.~Tutuc, Science {\bf 345}, 58 (2014).

\bibitem{HLZW}  B.~M. Hunt, J.~I.~A. Li, A.~A. Zibrov, L.~Wang, T.~Taniguchi, K.~Watanabe, 
J.~Hone, C.~R.~Dean, M.~Zaletel, R.~C. Ashoori, and A.~F. Young, Nat. Commun. {\bf 8}, 948 (2017). 

\bibitem{ZFJ}X. C. Zhang, D. R. Faulhaber, and H.W. Jiang, 
Phys. Rev. Lett. {\bf 95}, 216801 (2005).

\bibitem{BerryPh} M. V. Berry, Proc. R. Soc. London A {\bf 392}, 45 (1984).

\bibitem{WZ} F. Wilczek and  A. Zee, Phys. Rev. Lett. {\bf 52}, 2111 (1984).

\bibitem{GMP} S.~M.~Girvin, A.~H.~MacDonald, and P.~M.~Platzman,
Phys. Rev. B {\bf 33}, 2481 (1986).



\bibitem{ZLBF}  L. M. Zhang, Z. Q. Li, D. N. Basov, and M. M. Fogler, Z. Hao, and M. C. Martin, 
Phys. Rev. B {\bf 78}, 235408 (2008).

\bibitem{LHJ_asym} Z. Q. Li, E. A. Henriksen, Z. Jiang, Z. Hao, M. C. Martin, 
P. Kim, H. L. Stormer, andD. N. Basov, 
Phys. Rev. Lett. {\bf 102}, 037403 (2009).

\bibitem{ks_duality} K. Shizuya, Int. J. Mod. Phys. {\bf B} 31, 1750176 (2017).

\bibitem{JM} J. Jung and A. H. MacDonald,  Phys. Rev. B {\bf 89}, 035405 (2014). 

\bibitem{fn_one} Consider coupling to a scalar potential,
$H_{A} = -e\int d^{2}x A^{0} \rho$, and set
$A^{0}_{\bf p}\,  g^{10}_{\bf p}
=- c_{1} \ell\, (p_{x} - ip_{y})A^{0}_{\bf p}/\sqrt{2}
= -c_{1} \ell\, (E_{y} + iE_{x})_{\bf p}/\sqrt{2}$.


\bibitem{fnone} 
The spectra $(\hat{\epsilon}_{0_{\theta}}, \hat{\epsilon}_{1_{\theta}})$ acquire $O(X_{E}^{2/3})$ corrections 
in the very vicinities of $N_{\rm f} = N_{1}^{-}$ and $1+ N_{0}^{+}$. They do no harm to the spectra but formally make the electric dipole moment singular at those fillings  in the $E_{y}\rightarrow 0$ limit.


\bibitem{fntwo} 
Intuitively, the Coulombic energy $G^{nn}$ decreases with increasing $n$ 
because the electronic charge distribution $g^{nn}_{\bf p}$, 
with $g^{nn}_{\bf 0}=1$ and $g^{nm}_{{\bf p} \rightarrow 0} \propto p^{n-m}$, 
deviates more to ${\bf p}\not = 0$ in higher Landau levels.


\bibitem{KM83} 
M. Koshino and E. McCann, Phys. Rev. B {\bf 83}, 165443 (2011).

\bibitem{YRK} 
S. Yuan, R. Roldan and M. I. Katsnelson, Phys. Rev. B {\bf 84},
125455 (2011).

\bibitem{ZTM} 
F. Zhang, D. Tilahun, and A. H. MacDonald, Phys. Rev. B {\bf 85}, 165139 (2012).

\bibitem{ks_ABC} K. Shizuya, Phys. Rev. B {\bf 87}, 085413 (2013); 
Phys. Rev. B {\bf 89}, 165403 (2014).


\bibitem{BZZL} W.~Bao, Z.~Zhao, H.~Zhang, G.~Liu, P.~Kratz, L.~Jing, J.~Velasco,~Jr., 
D.~Smirnov, and C.~N.~Lau, Phys. Rev. Lett. {\bf 105}, 246601 (2010).

\bibitem{TWTJ} 
T. Taychatanapat, K. Watanabe, T. Taniguchi, and P. Jarilloo-Herrero,
Nat. Phys. {\bf 7}, 621 (2011).

\bibitem{LVTZ} 
Y. Lee, J. Velasco, Jr, D. Tran, F. Zhang, W. Bao, L. Jing, K. Myhro,
D. Smirnov, and C. N. Lau,
Nano. Lett. {\bf 13}, 1627 (2013).

\bibitem{EVT} 
H.~J.~van Elferen, A.~Veligura, N.~Tombros,  E.~V. Kurganova, B.~J.~van Wees, J.~C. Maan, 
 and U.~Zeitler, 
Phys. Rev. B {\bf 88}, 121302(R) (2013). 

\end{thebibliography}
\end{document}